\documentclass[preprint,prd,aps,showpacs,preprintnumbers,amsmath,amssymb]{revtex4-1}
\usepackage{graphics,epsfig,subfigure}
\usepackage{diagbox}
\usepackage[usenames]{color}
\usepackage{color}
\usepackage{graphicx}
\usepackage{amsfonts}
\usepackage{indentfirst}
\usepackage{booktabs}
\usepackage{hyperref}
\hypersetup{hypertex=ture,backref=true,colorlinks=ture,linkcolor=blue,anchorcolor=blue,citecolor=blue}
\usepackage{float}
\usepackage{multirow}

\begin{document}
\renewcommand{\baselinestretch}{1.3}
\newcommand\beq{\begin{equation}}
\newcommand\eeq{\end{equation}}
\newcommand\beqn{\begin{eqnarray}}
\newcommand\eeqn{\end{eqnarray}}
\newcommand\nn{\nonumber}
\newcommand\fc{\frac}
\newcommand\lt{\left}
\newcommand\rt{\right}
\newcommand\pt{\partial}

\title{\Large \bf Dirac-Proca stars}
\author{Tian-Xiang Ma, Chen Liang, Ji-Rong Ren, and Yong-Qiang Wang\footnote{yqwang@lzu.edu.cn, corresponding author
}
}

\affiliation{ $^{1}$Key Laboratory of Quantum Theory and Applications of MoE, Lanzhou Center for Theoretical Physics, Lanzhou University, Lanzhou 730000, China\\
	$^{2}$Key Laboratory of Theoretical Physics of Gansu Province, Institute of Theoretical Physics $\&$ Research Center of Gravitation, Lanzhou University, Lanzhou 730000, China\\
    $^{3}$School of Physical Science and Technology, Lanzhou University, Lanzhou 730000, China}

\begin{abstract}
  We consider a model consists of the Einstein gravity in four-dimensional spacetime, a Proca field and two Dirac fields through minimum coupling. By numerically solving this model, we obtain two types of solutions: synchronized frequency solutions and non-synchronized frequency solutions. We find that in the case of two kinds of matter fields, different families of solutions can be obtained in both synchronized and non-synchronized frequency cases, and the two families can shift to each other when certain parameters are changed. Moreover, we calculate the binding energy of multi-field solutions and give the relationship between binding energy $E$ and synchronized frequency $\tilde{\omega}$ (non-synchronized frequency $\tilde{\omega}_P$), so as to analyze the stability of the corresponding family.
\end{abstract}

\maketitle

\section{Introduction}\label{sec1}
Unlike fermions that make up ordinary matter and must satisfy the Pauli exclusion principle, bosons are particles that can occupy the same quantum state. This means that bosons can form very dense and compact objects that are held together by their own gravity and quantum pressure. Among these compact objects, the macroscopic Bose-Einstein condensate formed by bosons is called a boson star~\cite{Liebling:2012fv}. Generally speaking, they do not emit any electromagnetic radiation, nor do they have a surface or event horizon. Apart from the gravitational influence on the surrounding environment, they are basically invisible. It is precisely because of their characteristic that they are difficult to interact with ordinary matter except for gravity that boson stars have become one of the important candidates for dark matter~\cite{Sahni:1999qe,Hu:2000ke,Matos:2000ng,Lee:1995af,Pombo:2023ody}. So far, there is no direct evidence indicating the existence of boson stars, but there are still some methods that can be used to detect them. Even if boson stars have no observable effects in the electromagnetic aspect, they are still macroscopic objects formed by gravitational interactions, so they can be indirectly observed by the gravitational lensing effect that distorts the images of background light sources such as stars or galaxies~\cite{Cunha:2016bjh,Cunha:2017wao,Dabrowski:1998ac}. Moreover, with the rapid development of gravitational wave observation technology in recent decades, instruments such as LIGO and Virgo may be able to detect the characteristic signals of gravitational waves generated by boson stars when they merge with each other or with black holes~\cite{Yunes:2016jcc,Cardoso:2016oxy,Sennett:2017etc}, and the analysis of these gravitational wave signals also helps us to further understand the nature of boson stars.

After the classical boson star model, extended research was carried out in many different directions, one of which was to try to obtain different properties and mass upper bounds by changing the composition of the matter field. In addition to the classical boson star model composed of scalar fields~\cite{Kaup:1968zz,Colpi:1986ye,Siemonsen:2020hcg,Yoshida:1997qf,Ruffini:1969qy,Schunck:2003kk,Liebling:2012fv,Jetzer:1991jr}, there are also Dirac stars composed of spinor fields~\cite{Finster:1998ws,Dzhunushaliev:2018jhj,Bronnikov:2019nqa,Dzhunushaliev:2022wnd,Dolan:2015eua} (spin 1/2) and Proca stars composed of massive vector fields~\cite{Brito:2015pxa,Sanchis-Gual:2017bhw,Clough:2022ygm,DiGiovanni:2018bvo,Garcia-Saenz:2021uyv} (spin 1). Dirac stars are quantum entities consist of spinor fields, which to some extent represent the attempt of quantum gravity theory to unify gravity and quantum mechanics, and may thereby reveal some new aspects of the nature of spacetime. However, it is not particularly noteworthy in the field of astronomy. Proca stars consist of massive vector fields with spin 1, whose intrinsic angular momentum is one unit. They behave similarly to boson stars. Since Proca fields can also form stable structures, Proca stars have been widely studied in cosmology and astrophysics, such as for simulating black hole shadows~\cite{Herdeiro:2021lwl,Rosa:2022tfv}. They can also explain some gravitational wave events~\cite{Sanchis-Gual:2018oui,Tsukada:2020lgt,Sanchis-Gual:2022mkk}, such as GW190521. Some researchers believe that this event may be the merger of two Proca stars, rather than two black holes, because the initial mass of the objects was too large to be explained by stellar evolution~\cite{CalderonBustillo:2022dph,CalderonBustillo:2020fyi}.

Apart from changing the spin of particles that make up a boson star, there are also some boson star models composed of scalar fields in different states called multi-state bsoon star~\cite{Bernal:2009zy,Urena-Lopez:2010zva}. Following the study of multi-state boson stars, it was found that coupling gravity with multiple different fields can lead to soliton solutions that vary from those in the single-field case~\cite{Delgado:2020hwr,Blazquez-Salcedo:2019uqq,Dzhunushaliev:2019kiy,Sanchis-Gual:2021edp,Alcubierre:2022rgp,DiGiovanni:2021vlu,Cunha:2022tvk,Gervalle:2022fze,Sun:2022duv,Herdeiro:2021mol,Guerra:2019srj,Delgado:2020udb,Herdeiro:2023lze}, for example, Adding Yang-Mills fields~\cite{Dzhunushaliev:2019uft,Brihaye:2004nd,Straumann:1989tf}, Maxwell fields~\cite{Dzhunushaliev:2019kiy,Jetzer:1989av}, axion fields~\cite{Delgado:2020hwr,Guerra:2019srj,Delgado:2020udb}, Higgs bosons~\cite{Herdeiro:2023lze}, or superposing multiple same fields to form new configurations, such as star chains~\cite{Cunha:2022tvk,Gervalle:2022fze,Herdeiro:2021mol}, $\ell$-boson stars~\cite{Sanchis-Gual:2021edp}. This led to the development of multi-field boson star solutions obtained by coupling gravity with multiple fields of different spins, including solutions composed of Dirac fields and scalar fields~\cite{Liang:2022mjo} or scalar fields and Proca fields~\cite{Ma:2023vfa,Pombo:2023xkw}. So far, multi-field boson star solutions without scalar fields have not been thoroughly studied. However, considering Dirac-Proca star(DPS) model that uses Dirac fields and Proca fields coupled with gravity after excluding the scalar field components which make up traditional boson stars may have peculiar properties and stability conditions than single-state boson stars, and they may also have distinctive observational effects on dark matter and gravitational waves. Our work is to numerically solve the spherical symmetric model of two Dirac fields and one Proca field under the minimum coupling of Einstein gravity and obtain the properties of different families of solutions. All bosonic fields are in their ground state.

The structure of this paper is as follows. Sec.~\ref{sec2} introduces the basic framework and motion equations of the Einstein-Dirac-Proca system. In Sec.~\ref{sec3}, we derive the boundary conditions that each unknown function in the equation satisfies. In Sec.~\ref{sec4}, we present numerical results of the DPSs solutions and analyze their stability. In Sec.~\ref{sec5}, We present a summary and outlook.
\section{The model setup}\label{sec2}
The action of the minimal coupling of Einstein gravity with one Proca field and two Dirac fields is given by
\begin{equation}\label{equ1}
	S=\int \sqrt{-g} d^{4} x\left(\frac{R}{16 \pi G_0}+\mathcal{L}_{D}+\mathcal{L}_{P}\right),
\end{equation}
where $G_0$ is the gravitational constant, $R$ is the scalar curvature, the Lagrangian of the Dirac field and the Proca field, $\mathcal{L}_{D}$ and $\mathcal{L}_{P}$, their specific forms are
\begin{equation}\label{equ2}
  {\cal L}_{D}=-i\sum\limits_{k=1}^2 \left[ \frac{1}{2}\left(\hat{D}_\mu\overline{\Psi}^{(k)}\gamma^\mu\Psi^{(k)} - \overline{\Psi}^{(k)} \gamma^\mu\hat{D}_\mu\Psi^{(k)}\right) + \mu_D \overline{\Psi}^{(k)}\Psi^{(k)}\right]\,,
\end{equation}
\begin{equation}\label{equ3}
	\mathcal{L}_{P}=-\frac{1}{4} {\mathcal{F}}_{\alpha \beta} \overline{{\mathcal{F}}}^
  {\alpha \beta}-\frac{1}{2} \mu_P^{2} {\mathcal{A}}_{\alpha} \overline{\mathcal{A}}^{\alpha},
\end{equation}
where $\Psi^{(k)}$ are spinors with mass $\mu_D$ and $\mathcal{A}$ is Proca field, $\bar{\Psi}^{(k)}$ and $\bar{\mathcal{A}}$ are complex conjugates of their corresponding fields, $\mathcal{F}=d \mathcal{A}$.

The equations of motion derived from the action reads
\begin{equation}\label{equ4}
  R_{\alpha\beta} - \frac{1}{2}g_{\alpha\beta}R = 8\pi G_0 \left(T_{\alpha\beta(D)} + T_{\alpha\beta(P)}\right),
\end{equation}
\begin{equation}\label{equ5}
 \gamma^\mu\hat{D}_\mu\Psi^{(k)} - \mu_D \Psi^{(k)} = 0,
\end{equation}
\begin{equation}\label{equ6}
	\nabla_{\alpha} \mathcal{F}^{\alpha \beta}-\mu_{P}^{2} \mathcal{A}^{\beta}=0.
\end{equation}
where $T_{\alpha\beta}$ is the energy-stress tensor, which has the form
\begin{equation*}
  T_{\alpha\beta(D)} = \sum\limits_{k=1}^2 -\frac{i}{4}\left(\overline{\Psi}^{(k)}\gamma_\alpha\hat{D}_\beta\Psi^{(k)} + \overline{\Psi}^{(k)}\gamma_\beta\hat{D}_\alpha\Psi^{(k)} - \hat{D}_\alpha\overline{\Psi}^{(k)}\gamma_\beta\Psi^{(k)} - \hat{D}_\beta\overline{\Psi}^{(k)}\gamma_\alpha\Psi^{(k)}\right),
\end{equation*}
\begin{equation*}
	T_{\alpha \beta(P)}=\frac{1}{2}\left(\mathcal{F}_{\alpha \sigma} \overline{\mathcal{F}}_{\beta \gamma}+
	\overline{\mathcal{F}}_{\alpha \sigma} \mathcal{F}_{\beta \gamma}\right) g^{\sigma \gamma}-
	\frac{1}{4} g_{\alpha \beta} \mathcal{F}_{\sigma \tau} \overline{\mathcal{F}}^{\sigma \tau}+
	\frac{1}{2} \mu_P^{2}\left[\mathcal{A}_{\alpha} \overline{\mathcal{A}}_{\beta}+
	\overline{\mathcal{A}}_{\alpha} \mathcal{A}_{\beta}-g_{\alpha \beta} \mathcal{\mathcal { A }}
	_{\sigma} \overline{\mathcal{A}}^{\sigma}\right].
\end{equation*}

The action~(\Ref{equ1}) is invariant under global $U(1)$ transformations $\Psi^{(k)}\rightarrow e^{i\alpha}\Psi^{(k)}$ and $\mathcal{A}^{\beta}\rightarrow e^{i\alpha}\mathcal{A}^{\beta}$ with a constant $\alpha$, which means that there is a conserved current $J^\alpha$ for this system.
\begin{equation}\label{equ7}
  J_D^{\alpha} = \overline{\Psi}\gamma^\alpha\Psi,
\end{equation}
\begin{equation}\label{equ8}
  J_P^{\alpha} = \frac{i}{2}\left[\overline{\mathcal{F}}^{\alpha \beta} \mathcal{A}_
	{\beta}-\mathcal{F}^{\alpha \beta} \overline{\mathcal{A}}_{\beta}\right].
\end{equation}
Integrating the time component of this four-dimensional current over a spacelike hypersurface $\varSigma$ gives the conserved Noether charge $Q$
\begin{equation}\label{equ9}
  Q_D = \int_{\varSigma}J_D^t, \qquad Q_P = \int_{\varSigma}J_P^t,
\end{equation}
This conserved charge corresponds to the particle number of the respective matter fields.

To obtain spherically symmetric solutions, we use the following form of spherically symmetric metric:
\begin{equation}\label{equ10}
  ds^2 = -N(r)\sigma^2(r)dt^2 + \frac{dr^2}{N(r)} + r^2\left(d\theta^2 + \sin^2\theta d\varphi^2\right)\,,
\end{equation}
where $N(r) = 1 - {2m(r)}/{r}$. The two pairs of Dirac fields are given by~\cite{Herdeiro:2017fhv}:
\begin{equation}\label{equ11}
   \Psi^{(1)} = \begin{pmatrix}\cos(\frac{\theta}{2})[(1 + i)f(r) - (1 - i)g(r)]\\ i\sin(\frac{\theta}{2})[(1 - i)f(r) - (1 + i)g(r)]\\-i\cos(\frac{\theta}{2})[(1 - i)f(r) - (1 + i)g(r)]\\ -\sin(\frac{\theta}{2})[(1 + i)f(r) - (1 - i)g(r)] \end{pmatrix}e^{i\frac{\varphi}{2} - i\omega_D t}\,,
\end{equation}
\begin{equation}\label{equ12}
  \Psi^{(2)} = \begin{pmatrix}i\sin(\frac{\theta}{2})[(1 + i)f(r) - (1 - i)g(r)]\\ \cos(\frac{\theta}{2})[(1 - i)f(r) - (1 + i)g(r)]\\ \sin(\frac{\theta}{2})[(1 - i)f(r) - (1 + i)g_n(r)]\\ i\cos(\frac{\theta}{2})[(1 + i)f(r) - (1 - i)g(r)] \end{pmatrix}e^{-i\frac{\varphi}{2} - i\omega_D t}\,,
\end{equation}
and the Proca field is
\begin{equation}\label{equ13}
	\mathcal{A}=[F(r) d t+i G(r) d r] e^{-i \omega_P t}
\end{equation}
Substituting the above ansatz into the field equations ~(\ref{equ4}--\ref{equ6}) yields the following system of ordinary differential equations:
\begin{equation}\label{equ14}
  f^\prime + \left(\frac{N^\prime}{4N} + \frac{\sigma^\prime}{2\sigma} + \frac{1}{r\sqrt{N}} + \frac{1}{r}\right)f + \left(\frac{\mu}{\sqrt{N}} - \frac{\omega_D}{N\sigma}\right)g = 0 \,,
\end{equation}
\begin{equation}\label{equ15}
  g^\prime + \left(\frac{N^\prime}{4N} + \frac{\sigma^\prime}{2\sigma} - \frac{1}{r\sqrt{N}} + \frac{1}{r}\right)g + \left(\frac{\mu}{\sqrt{N}} + \frac{\omega_D}{N\sigma}\right)f = 0\,,
\end{equation}
\begin{equation}\label{equ16}
	\frac{d}{d r}\left\{\frac{r^{2}\left[F^{\prime}-\omega_P G\right]}{\sigma}\right\}=\frac{\mu_P^{2} r^{2} F}{\sigma N},
\end{equation}
\begin{equation}\label{equ17}
	\omega_P G-F^{\prime}=\frac{\mu_P^{2} \sigma^{2} N G}{\omega_P},
\end{equation}
\begin{equation}\label{equ18}
  m^\prime = \frac{32\pi G_0 r^2 \omega_D(f^2+g^2)}{\sqrt{N}\sigma} + 4 \pi G_0 r^2 \left[\frac{(F^{\prime}-\omega_P G)^2}{2\sigma^2}+\frac{\mu_P^2}2\left(G^2N+\frac{F^2}{N\sigma^2}\right)\right] ,
 \end{equation}
 \begin{equation}\label{equ19}
  \frac{\sigma^\prime}{\sigma} = \frac{32 \pi G_0 r}{\sqrt{N}}\left[gf^{\prime}-fg^{\prime}+\frac{\omega_D(f^2+g^2)}{N\sigma}\right]+4 \pi G_0 r \mu_P^2\left(G^2+\frac{F^2}{N^2\sigma^2}\right),
\end{equation}
The specific form of the Noether charge is:
\begin{equation}\label{equ20}
 Q_D = 16\pi\int_0^\infty r^2\frac{f^2 + g^2}{\sqrt{N}}dr, \quad Q_P = 4\pi\int_0^\infty r^2\frac{\left(\omega_P G-F^{\prime}\right) G}{\sigma}.
\end{equation}

\section{Boundary conditions}\label{sec3}

To solve the system of ordinary differential equations derived in the previous section, we have to specify the boundary conditions for each unknown function. Since they are solutions with asymptotic flatness, the metric functions $m(r)$ and $\sigma(r)$ have to satisfy the following boundary conditions:
\begin{equation}
m(0) = 0,\qquad \sigma(0) = \sigma_0,\qquad m(\infty) = M,\qquad \sigma(\infty) = 1,
\end{equation}
where the ADM mass $M$ and $\sigma_0$ are unknown constants. In addition, the matter field should vanish at infinity:
\begin{equation}
  f(\infty) = 0,\qquad g(\infty) = 0\,,\quad F(\infty) = 0, \qquad G(\infty) = 0.
\end{equation}
Expanding equations~(\ref{equ14}--\ref{equ17}) near the origin, we obtain that the field functions satisfy the following condition at the origin:
\begin{equation}
   f(0) = 0,\qquad \left.\frac{dg(r)}{dr}\right|_{r = 0} = 0,\qquad \left.\frac{dF(r)}{dr}\right|_{r = 0} = 0, \qquad G(0) = 0.
\end{equation}

\section{Numerical results}\label{sec4}

To facilitate numerical calculations, we use dimensionless quantities:
\begin{equation}\label{equ22}
  \begin{split}
  \tilde{r} \rightarrow r/\rho, \quad f \rightarrow \frac{\sqrt{4\pi\rho}}{M_{Pl}}f,\quad g \rightarrow \frac{\sqrt{4\pi\rho}}{M_{Pl}}g,
  \quad \tilde{\omega}_D \rightarrow \omega_D\rho, \quad \tilde{\mu}_D \rightarrow \mu_D\rho,
  \\\quad \tilde{F} \rightarrow \frac{\sqrt{4\pi}}{M_{Pl}}F, \quad \tilde{G} \rightarrow \frac{\sqrt{4\pi}}{M_{Pl}}G,
  \quad \tilde{\omega}_P \rightarrow \omega_P\rho, \quad \tilde{\mu}_P \rightarrow \mu_P\rho,
  \end{split}
\end{equation}
where $M_{Pl} = 1/\sqrt{G_0}$ is the Planck mass. $\rho$ is a positive constant whose dimension is length, we let the
constant $\rho$ be $1/\mu_S$. Additionally, we introduce a new radial variable $x$:
\begin{equation}
x = \frac{\tilde{r}}{1+\tilde{r}},
\end{equation}
where the radial coordinate $\tilde{r}\in[0,\infty)$, so $x\in[0,1]$. We utilize the finite element method to numerically solve the system of differential equations. The integration region $0\le x\le 1$ is discretized into 1000 grid points. The Newton-Raphson method is employed as our iterative approach. To ensure the accuracy of the computational results, we enforce a relative error criterion of less than $10^{-5}$.

To ensure the correctness of our numerical calculations, we need to check the numerical precision by validating physical constraints~\cite{Herdeiro:2021teo,Herdeiro:2022ids}, besides using the numerical analysis methods mentioned before. In this study, we compared the asymptotic mass and the Komar mass of the electron in the numerical solution, and found that the difference between them was always less than $10^{-5}$.

In our model, the ground state Dirac field functions $f$ and $g$ have no nodes in the radial direction, so they are denoted by $D_0$ where the subscript is the total number of radial nodes of the field functions. The ground state field functions $F$ and $G$ of Proca field have one node in the radial direction, so they are denoted by $P_1$, where the subscript has the same meaning as that of the scalar field. Therefore, in this model we use $D_0P_1$ to represent the coexisting state of the Dirac field and Proca field. Next, we  analyze the different classes of families of solutions of Dirac-Proca stars in detail.
\subsection{ Synchronized frequency }
We found through the analysis of numerical calculations that by changing the Proca field mass $\tilde{\mu}_P$ in the synchronized frequency solutions $(\tilde{\omega}=\tilde{\omega}_D=\tilde{\omega}_P)$, different DPS solutions can be obtained. According to the different behaviors of the solutions at the extreme value of the synchronized frequency $\tilde{\omega}$, we can divide the synchronized frequency solutions into two categories: \textit{D-P} and \textit{P-P}. For $0.828\leq \tilde{\mu}_P \leq 0.856$, the DPS solutions belong to the \textit{P-P} family; for $0.857 < \tilde{\mu}_P \leq 0.985$, the DPS solutions belong to the \textit{D-P} family. Next, we  discuss in detail the properties of these two types of solutions.
\subsubsection{P-P family}
\begin{figure}[!htbp]
\begin{center}
  \includegraphics[height=.26\textheight]{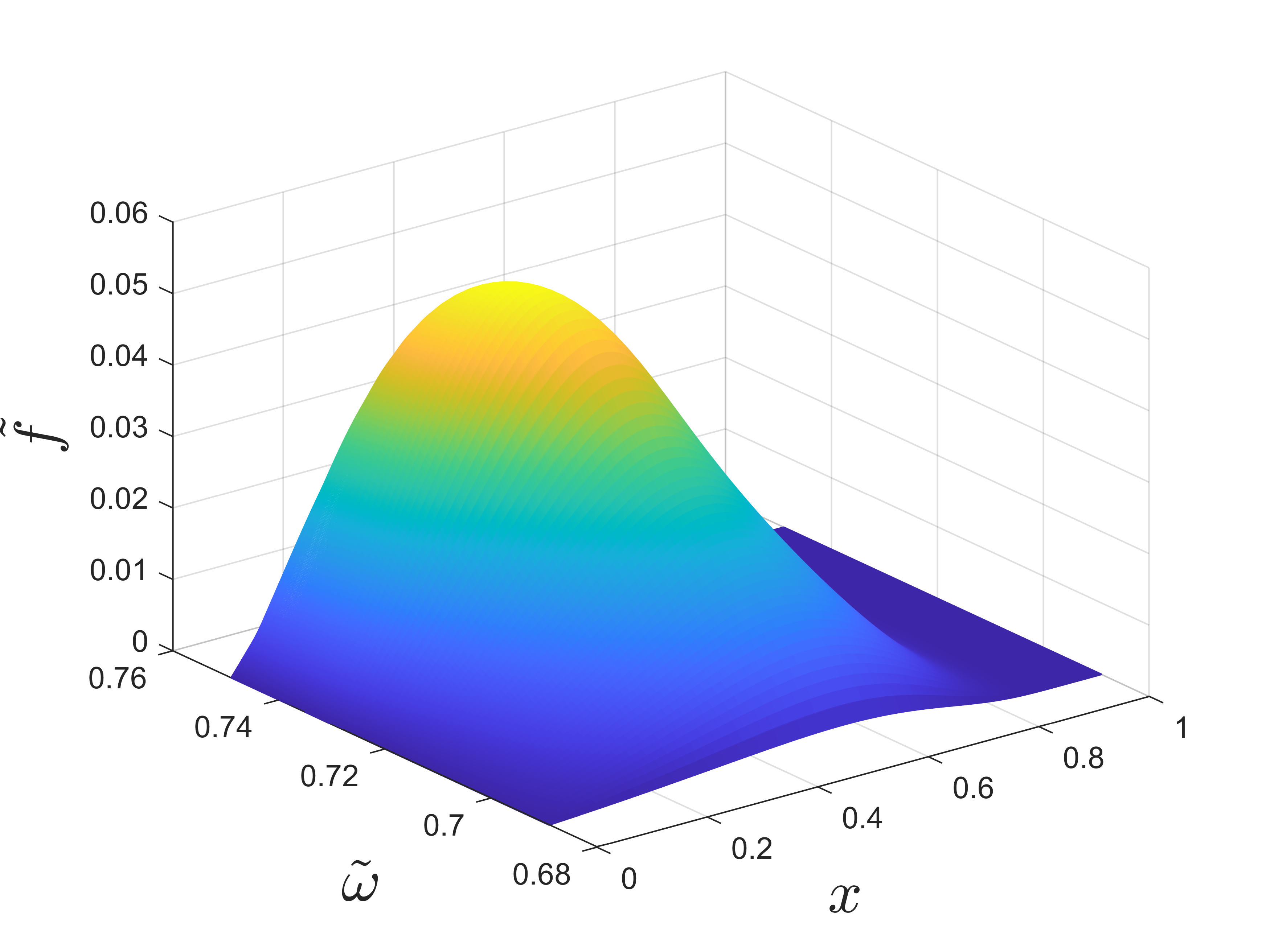}
  \includegraphics[height=.26\textheight]{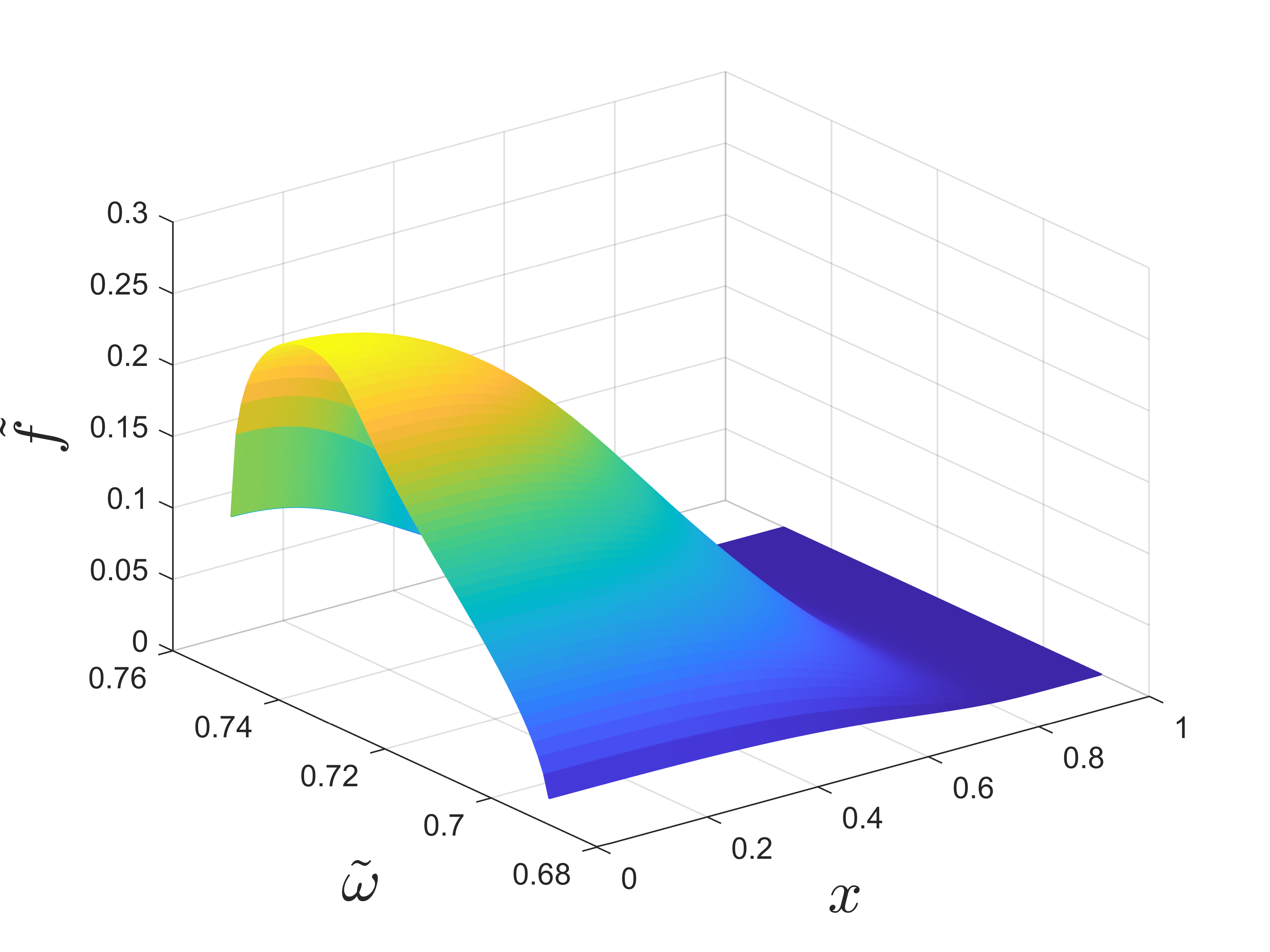}
  \includegraphics[height=.26\textheight]{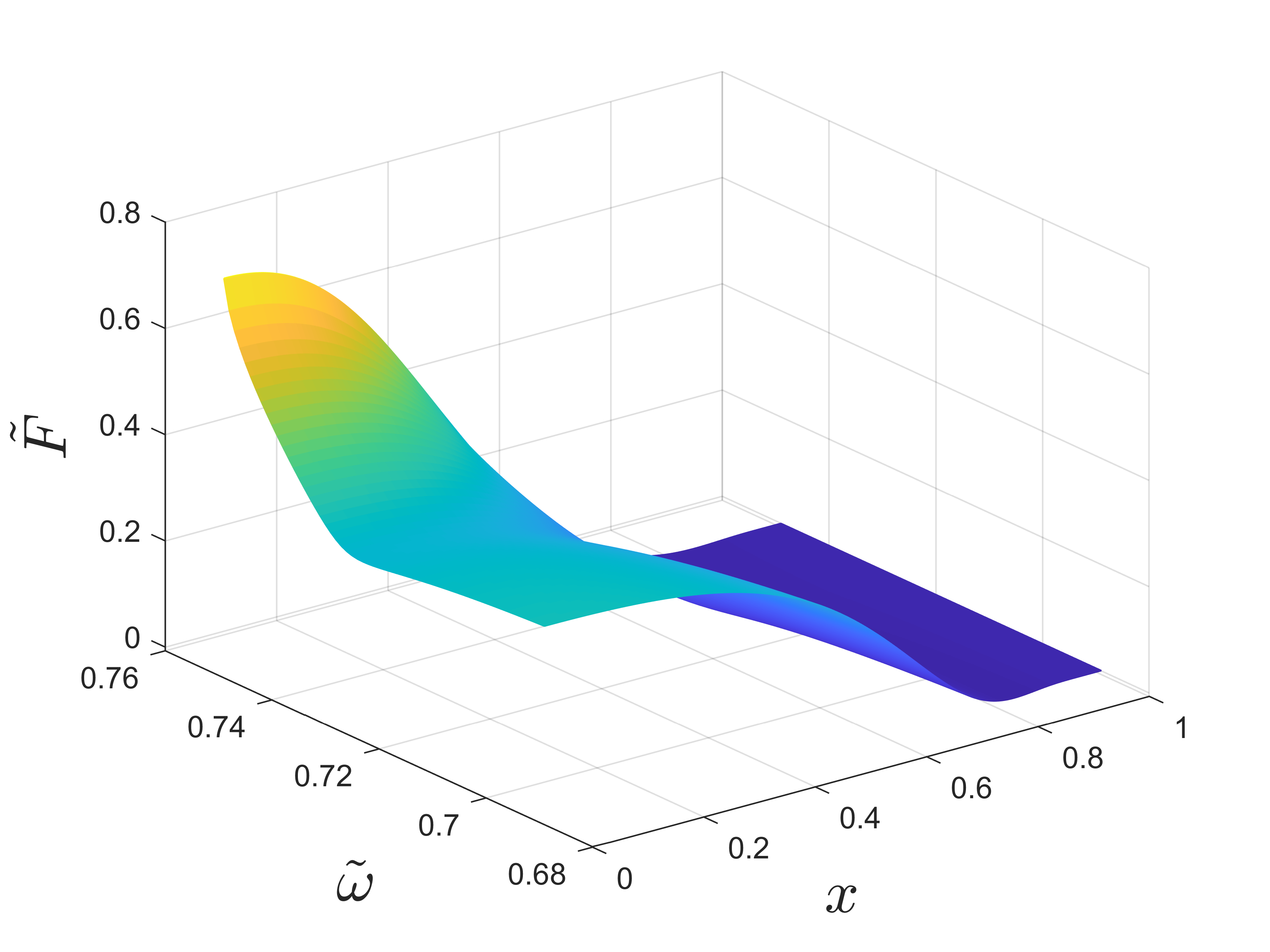}
  \includegraphics[height=.26\textheight]{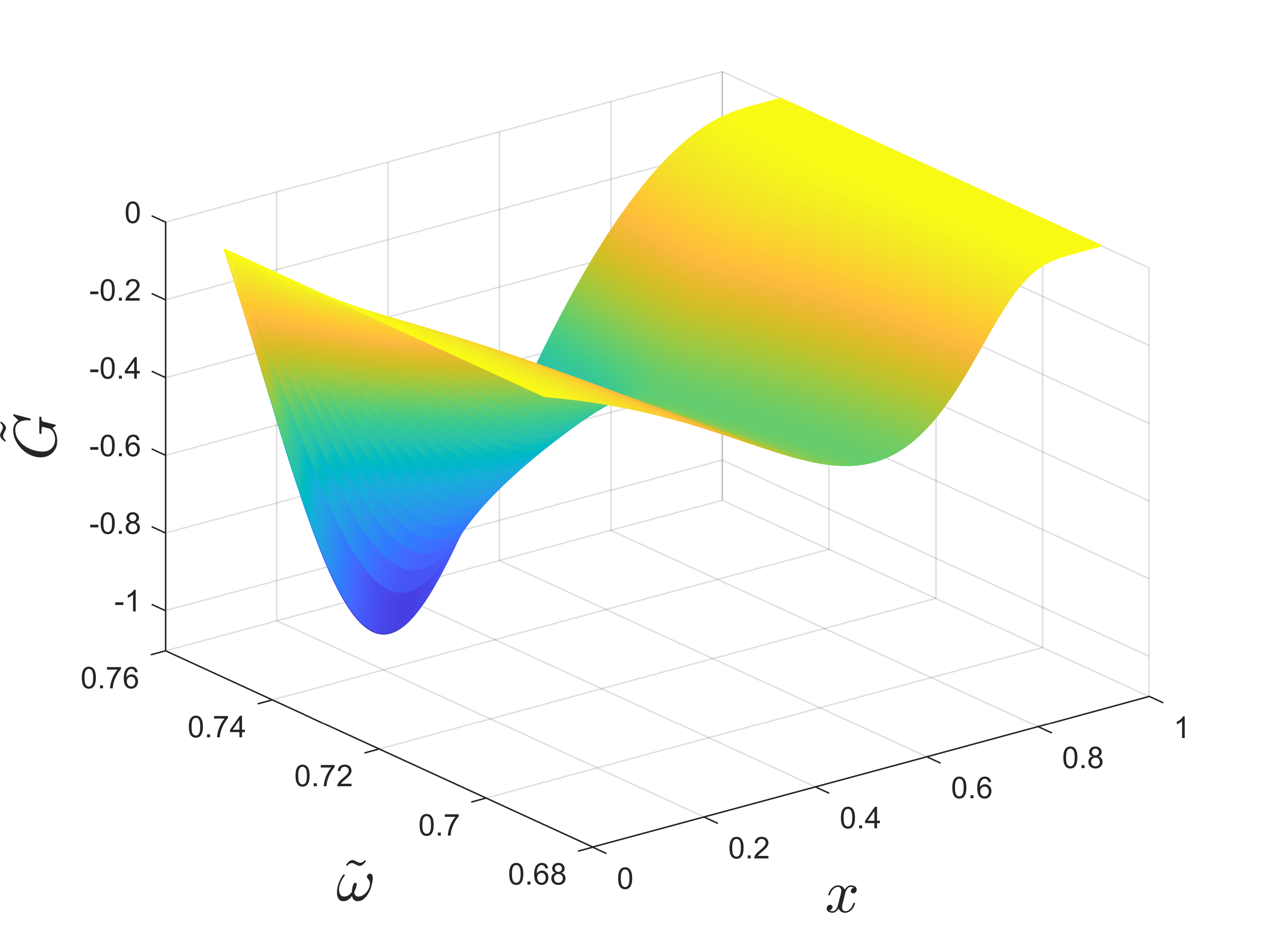}
\end{center}
\caption{The matter field functions $\tilde{f}$, $\tilde{g}$, $\tilde{F}$ and $\tilde{G}$ as functions of $x$ and $\tilde{\omega}$ for $\tilde{\mu}_P = 0.845$. All of them are in the first branch.}
\label{field1}
\end{figure}

\begin{figure}[!htbp]
\begin{center}
    \includegraphics[height=.26\textheight]{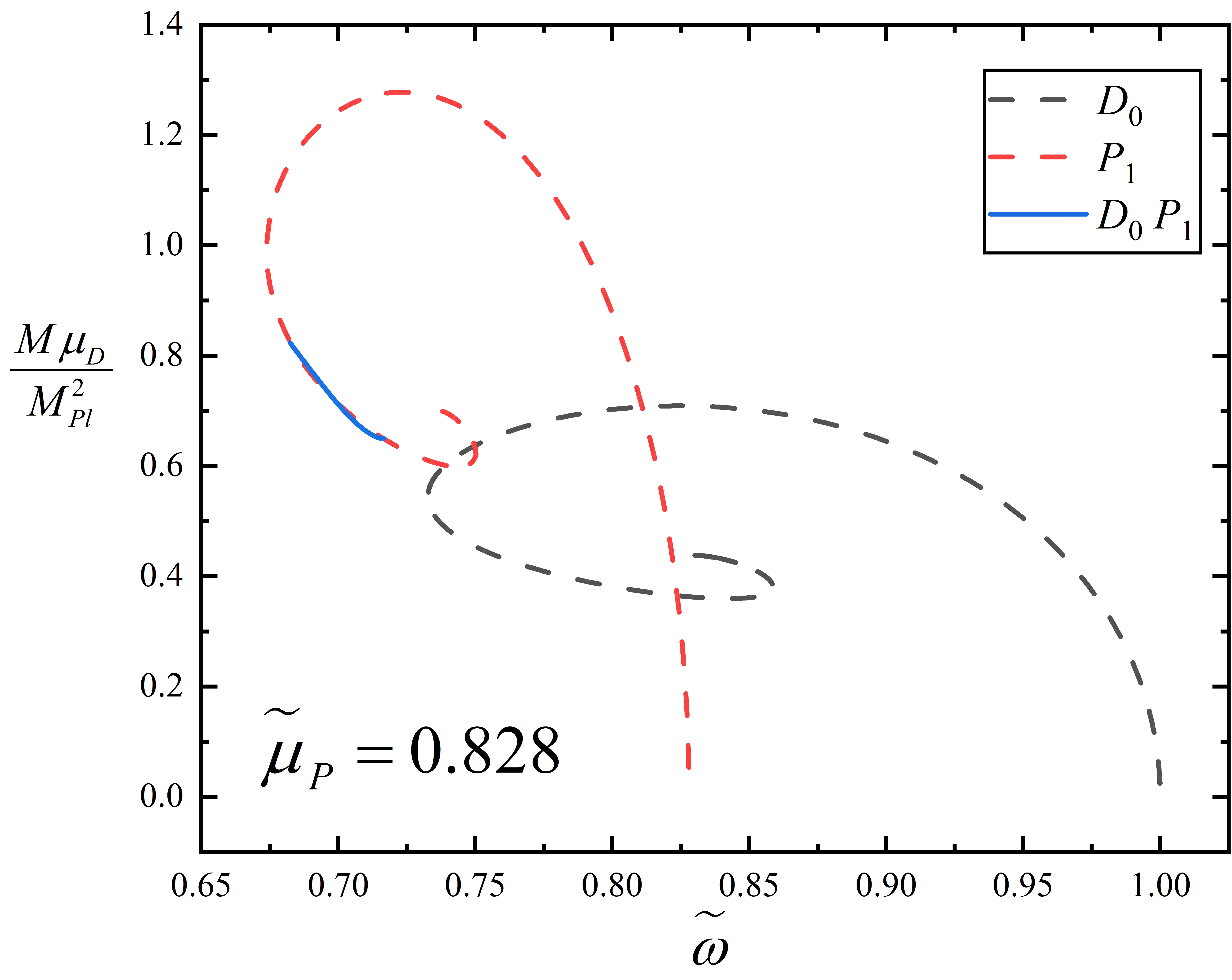}
    \includegraphics[height=.26\textheight]{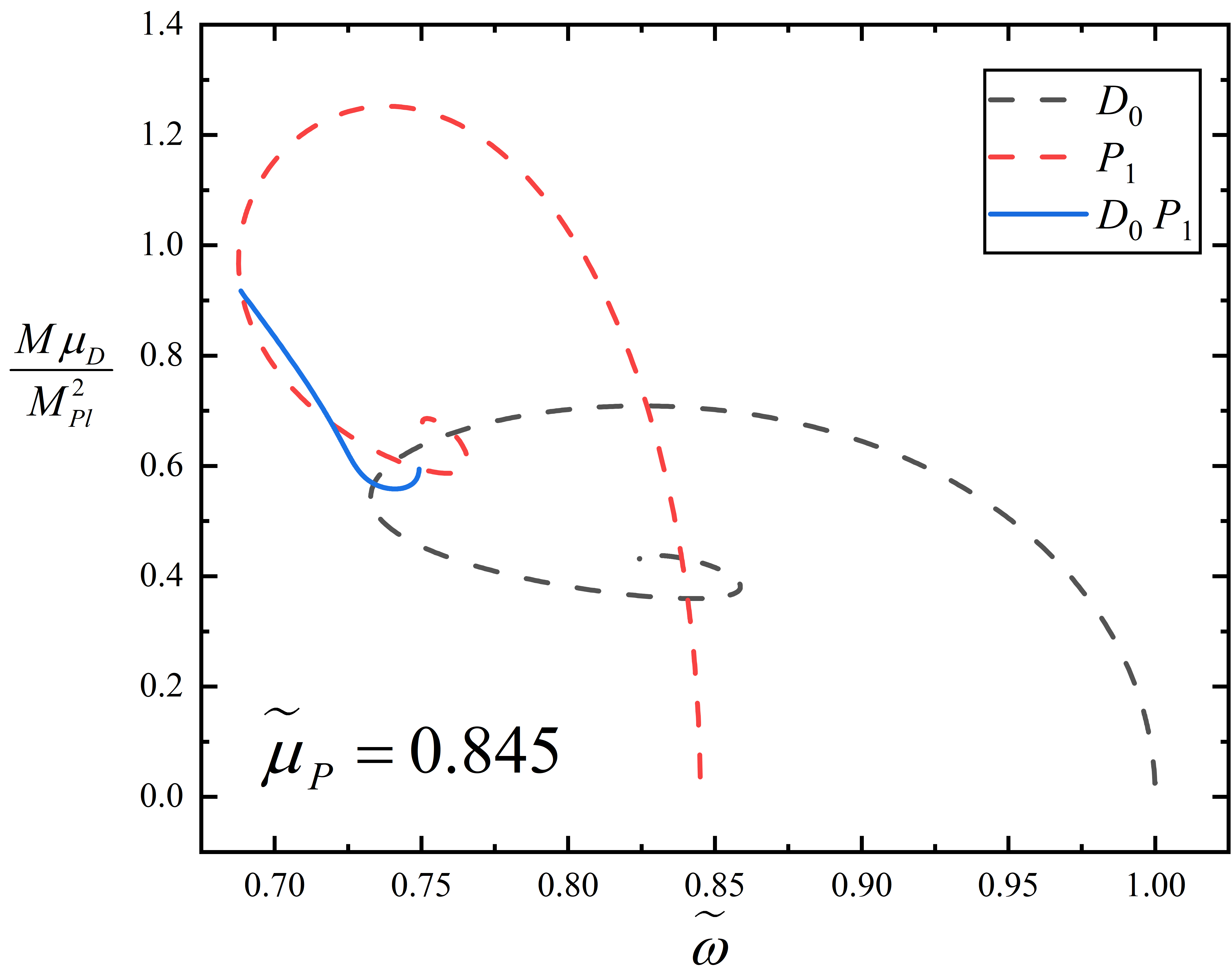}
    \includegraphics[height=.26\textheight]{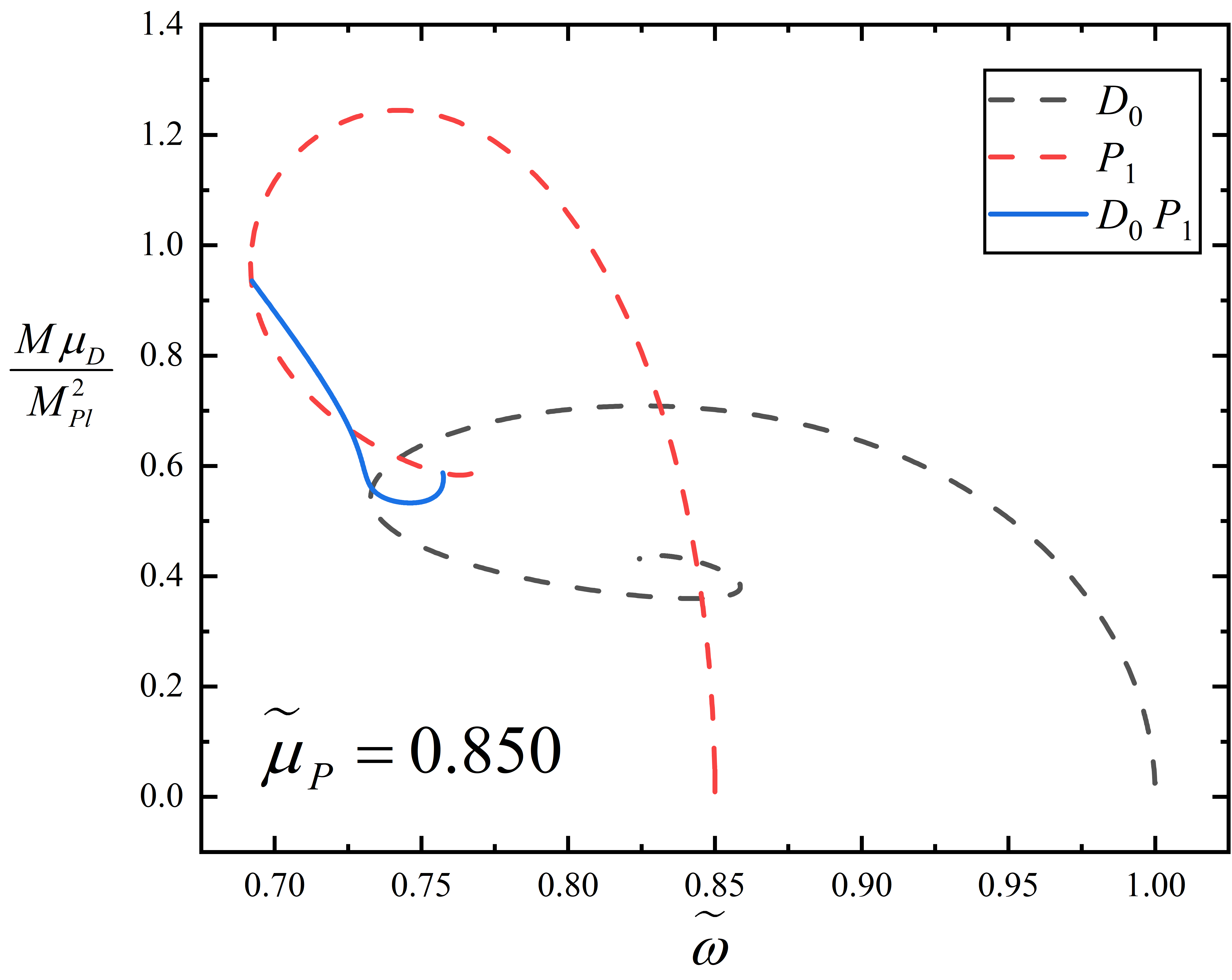}
    \includegraphics[height=.26\textheight]{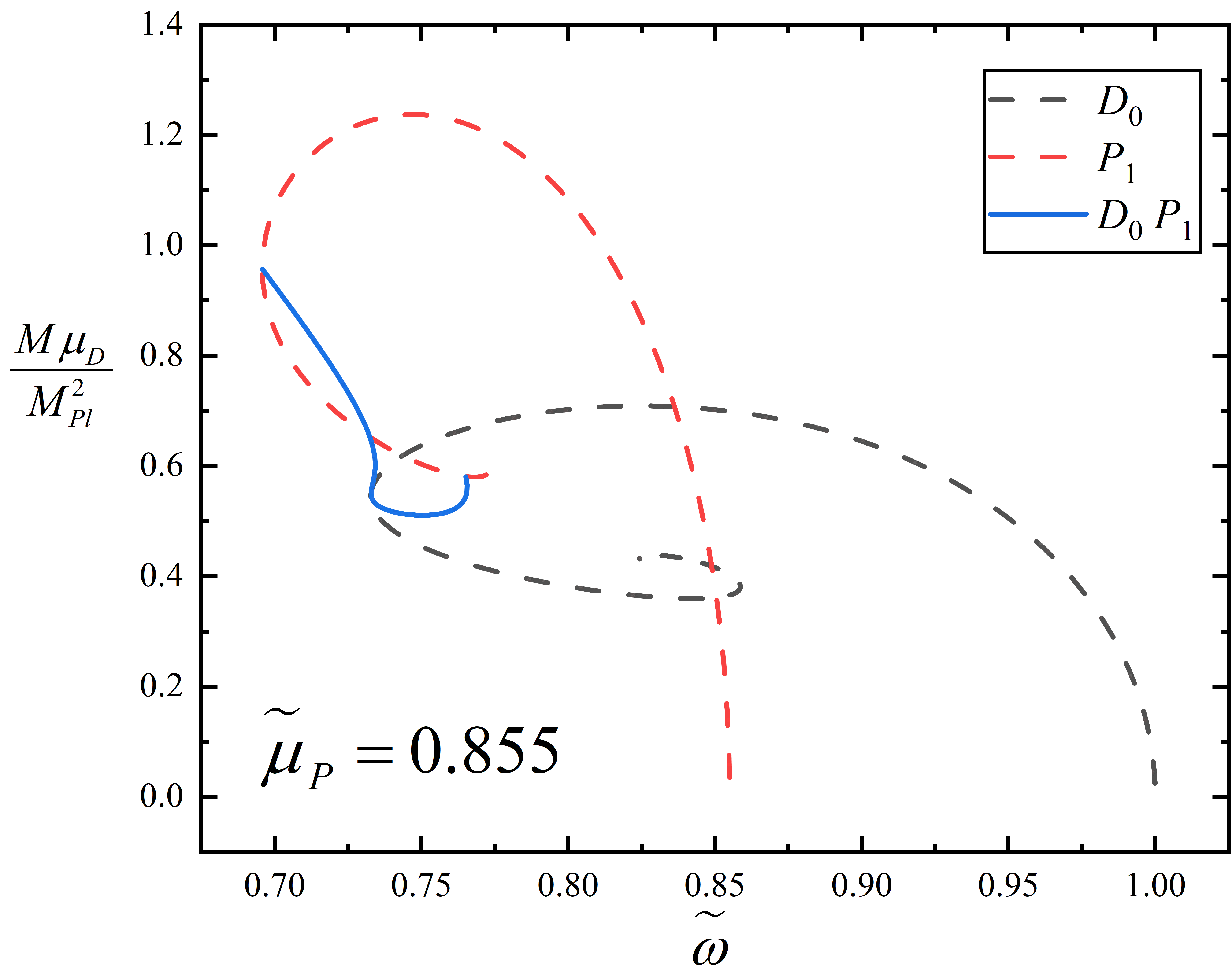}
\end{center}
\caption{ The ADM mass $M$ of the DPSs as a function of the synchronized frequency $\tilde{\omega}$ for several values of $\tilde{\mu}_P$.
}
\label{ADM1}
\end{figure}

The first type of synchronized frequency solutions can be obtained when the value of $\tilde{\mu}_P$ is small. Since the two endpoints of the solution's ADM curve fall on the single-field curve of the Proca field, we call it a \textit{P-P} solution. When the synchronized frequency $\tilde{\omega}$ takes the minimum or maximum value, the Dirac field disappears, and the multi-field solution becomes a single-field solution with only Proca field components. Fig.~\ref{field1} shows the field functions at different synchronized frequencies $\tilde{\omega}$. The top two graphs are Dirac field functions ($f$ and $g$), and the bottom two graphs are Proca field functions ($F$ and $G$). It can be seen that except for the image of $F$ having one node, all other field functions do not have nodes. It is easy to see from the Fig.~\ref{field1} that the peak absolute value of the Dirac field function in the \textit{P-P} family does not monotonically change with $\tilde{\omega}$, $\left\lvert f_{max}\right\rvert $ and $\left\lvert g_{max}\right\rvert $ both exhibit a trend of increasing first and then decreasing as $\tilde{\omega}$ increases. For Proca field, $\left\lvert F_{max}\right\rvert $ and $\left\lvert G_{max}\right\rvert $ increase monotonically with $\tilde{\omega}$.

Next, let's analyze the ADM mass of the multi-field solution. By changing the value of $\tilde{\mu}_P$ within a certain range, we can obtain different \textit{P-P} solutions. The relationship between their ADM mass and the synchronized frequency $\tilde{\omega}$ is shown in Fig.~\ref{ADM1}. The black dashed line represents the ground state Dirac field $(D_0)$, the red dashed line represents the Proca field $(P_1)$, and the blue solid line represents the multi-field solutions $(D_0 P_1)$. We can see from this that the \textit{P-P} solutions have the characteristic that both ends of the multi-field solutions fall on the Proca field curve. Their shape is like a spoon. With increasing $\tilde{\mu}_P$, the existence domain of the multi-field solutions gradually increases, and there will be a certain degree of distortion at the connection between the "spoon head" and the "spoon handle". In this domain, the ADM mass changes from a single-valued function of $\tilde{\omega}$ to a multi-valued function. The curve starts from the Proca field and gradually extends to the right. At this time, the proportion of Dirac field component increases gradually from 0 and then decreases gradually. Finally, it terminates at another position on the Proca field, and the Dirac field disappears again.

\begin{figure}[!htbp]
  \begin{center}
      \includegraphics[height=.26\textheight]{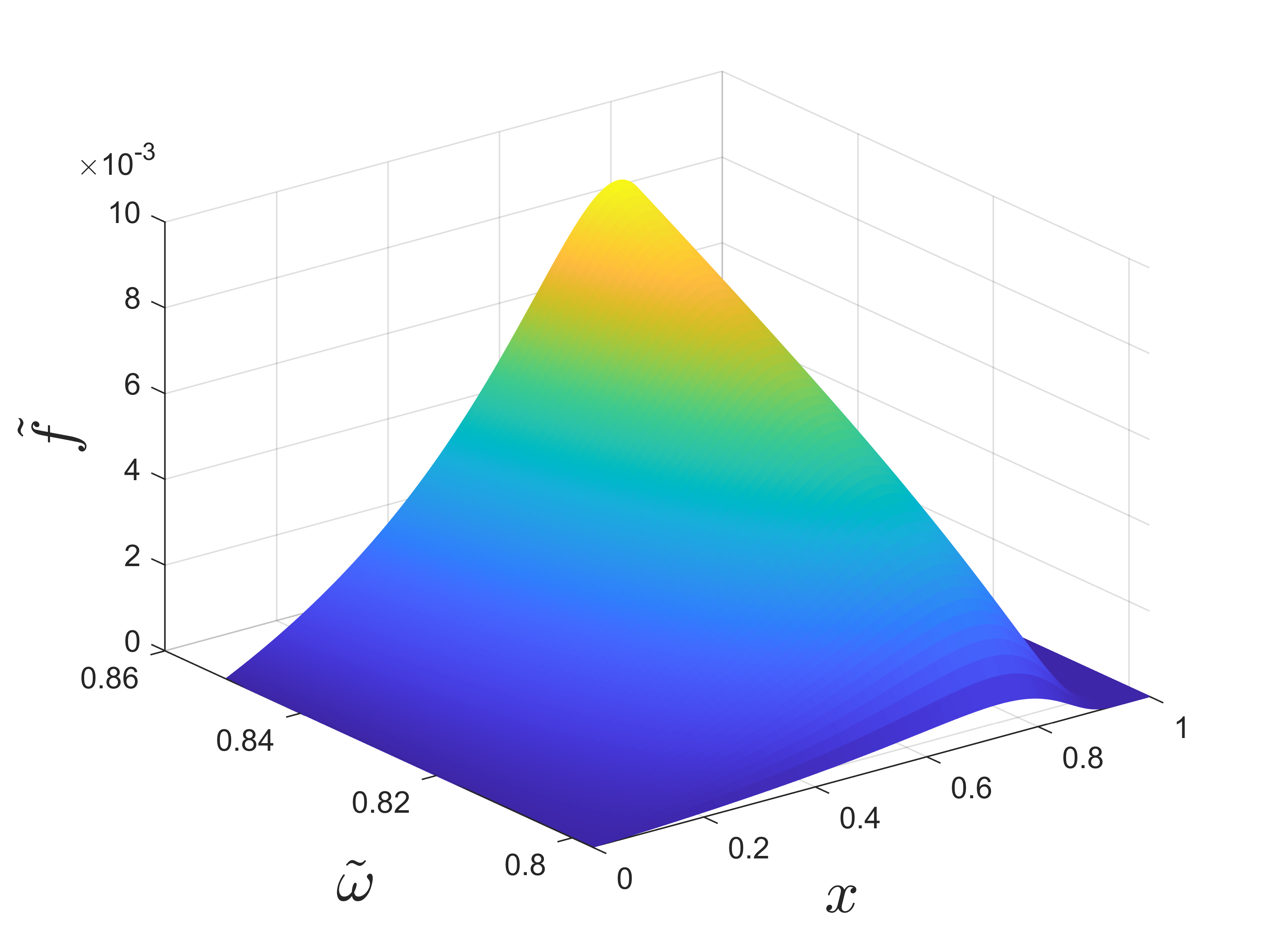}
      \includegraphics[height=.26\textheight]{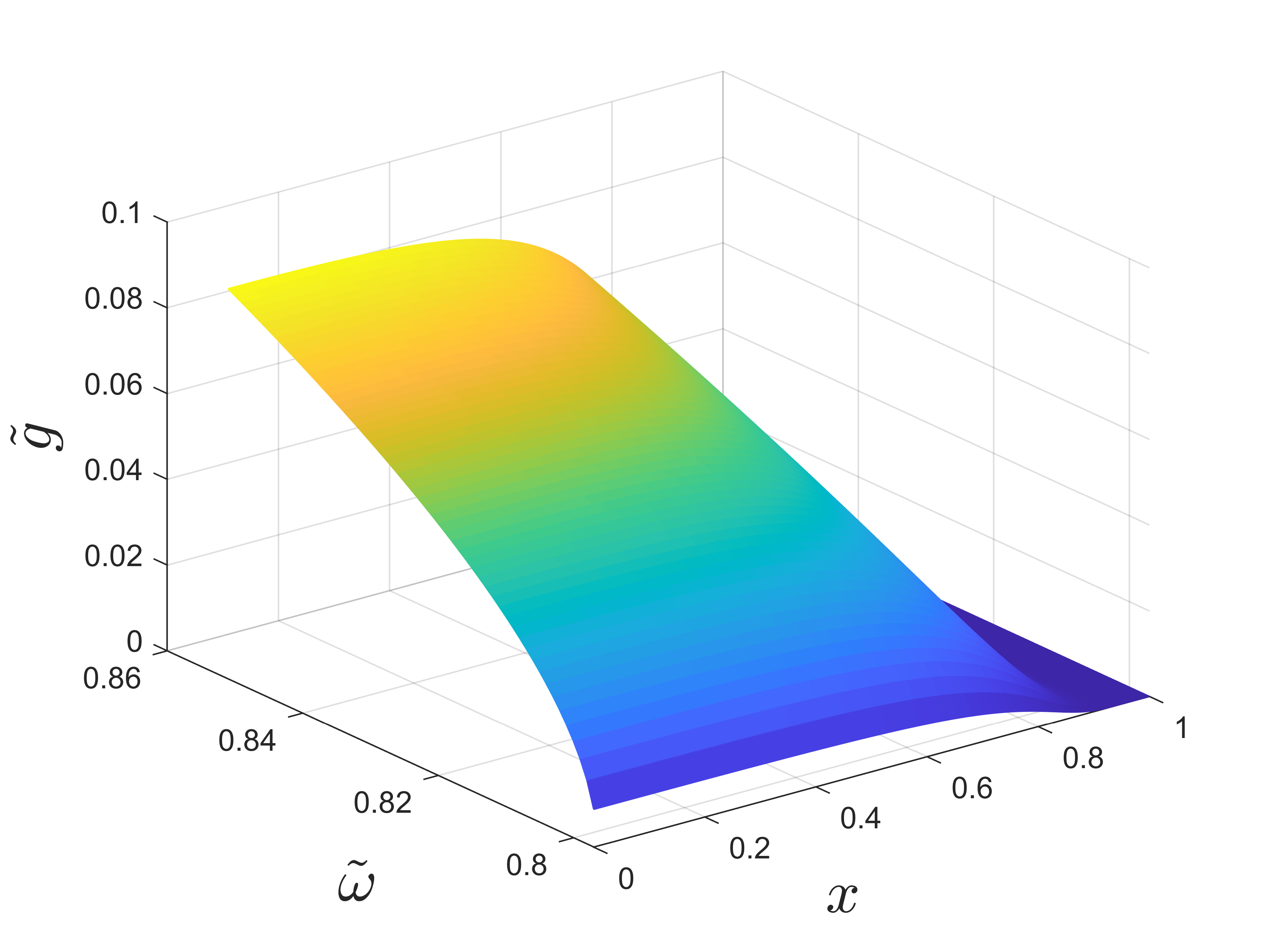}
      \includegraphics[height=.26\textheight]{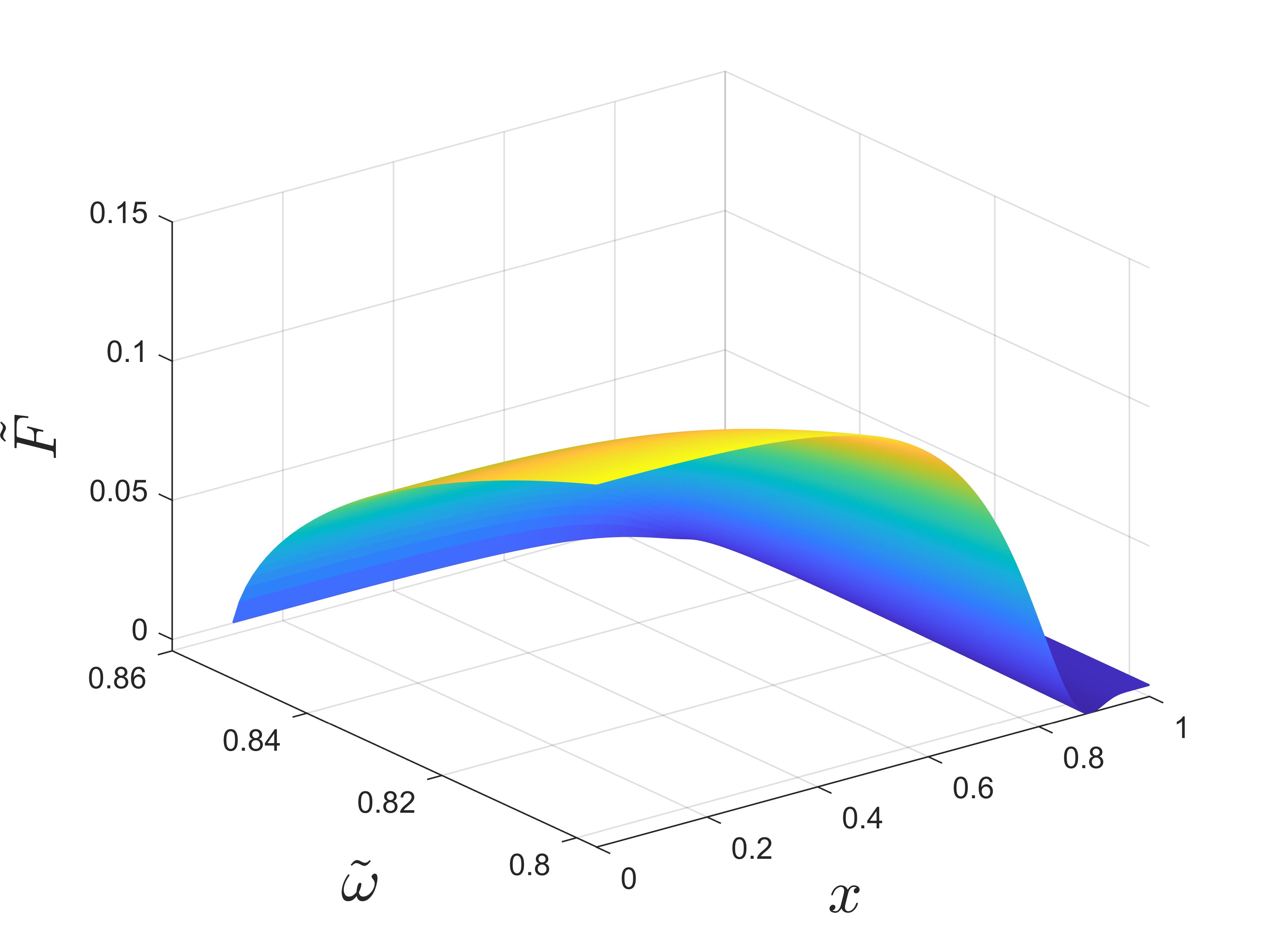}
      \includegraphics[height=.26\textheight]{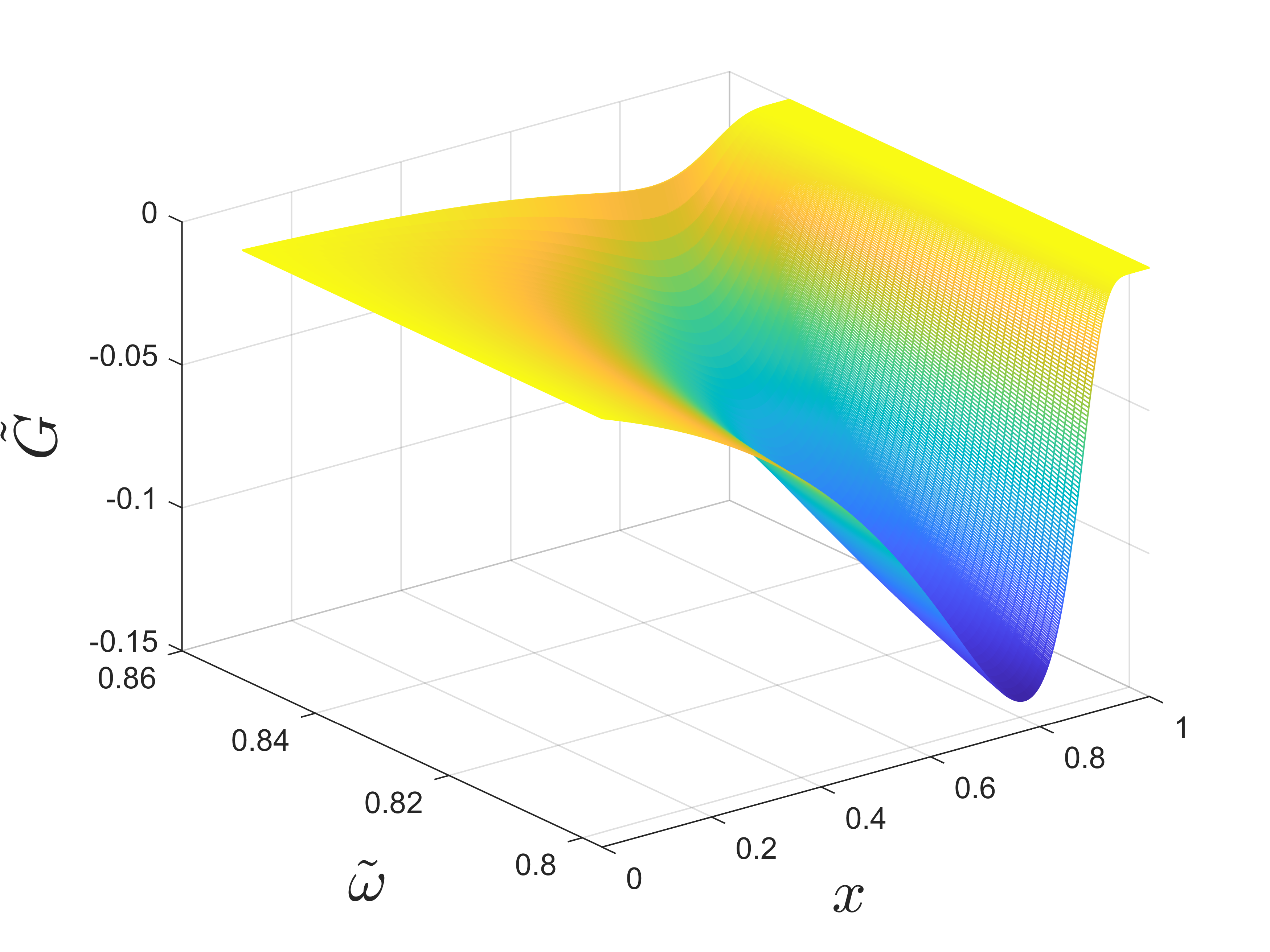}
  \end{center}
  \caption{The matter functions $\tilde{f}$, $\tilde{g}$, $\tilde{F}$ and $\tilde{G}$ as functions of $x$ and $\tilde{\omega}$ for $\tilde{\mu}_P = 0.93$.}
  \label{field2}
\end{figure}

\begin{figure}[!htbp]
  \begin{center}
      \includegraphics[height=.26\textheight]{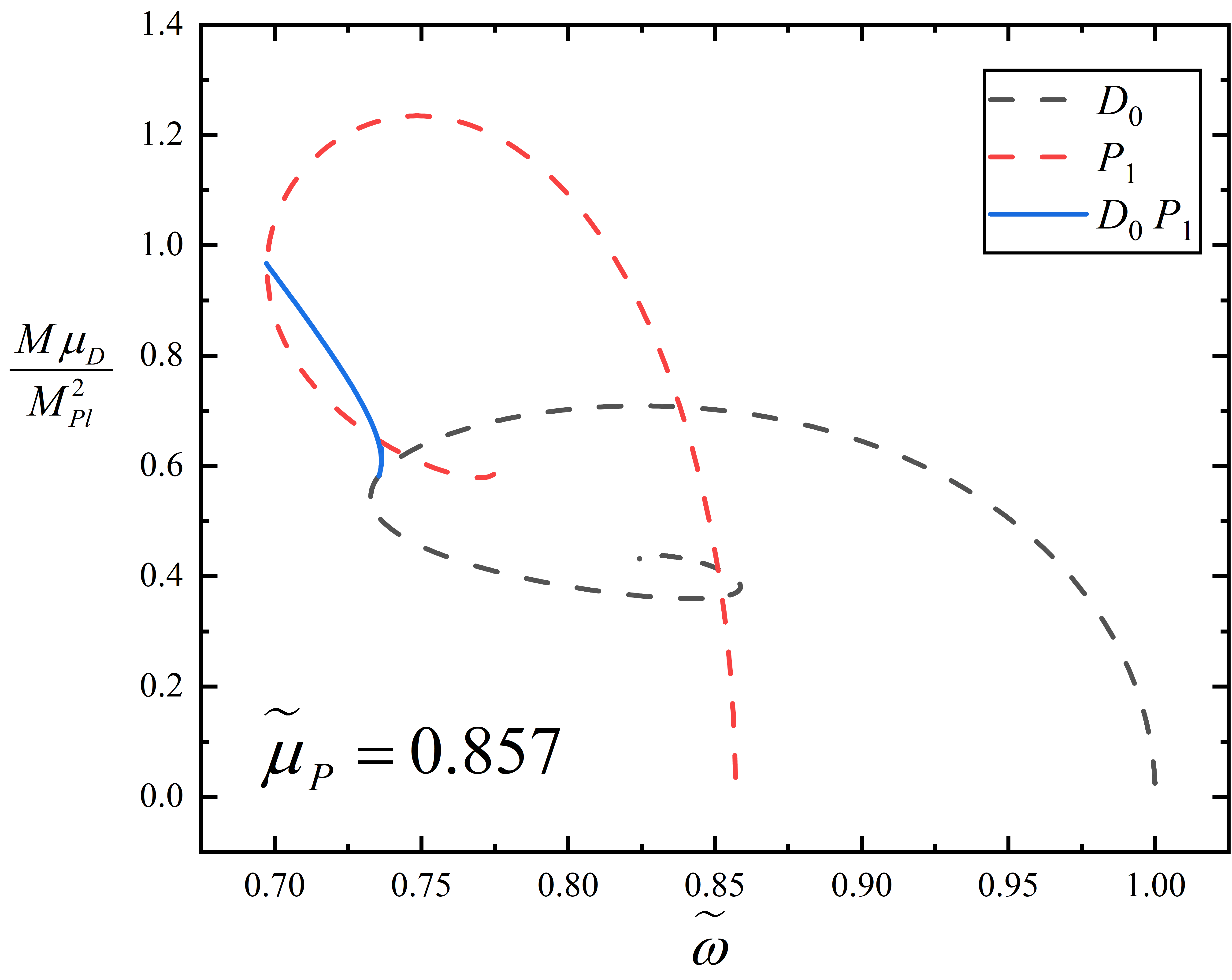}
      \includegraphics[height=.26\textheight]{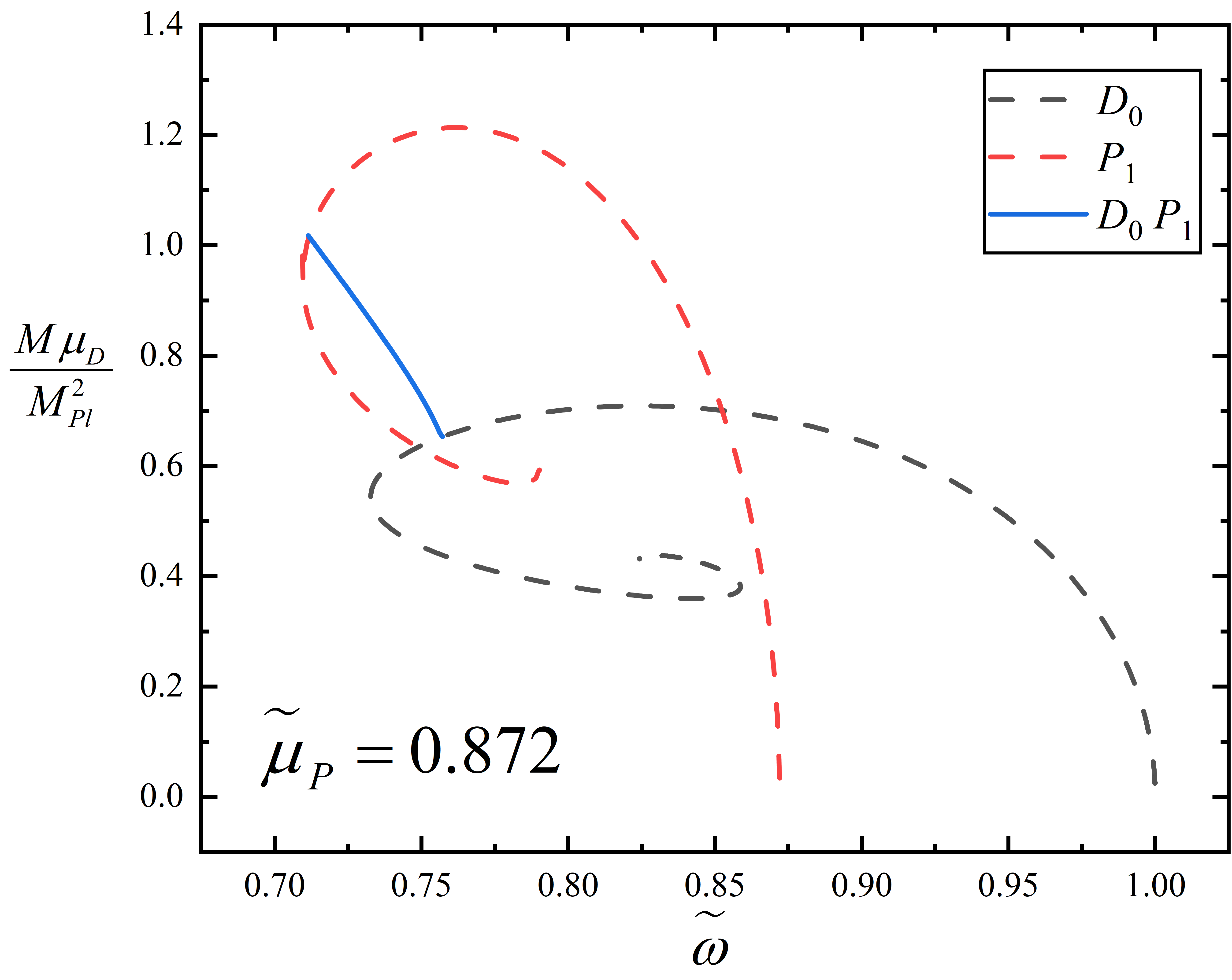}
      \includegraphics[height=.26\textheight]{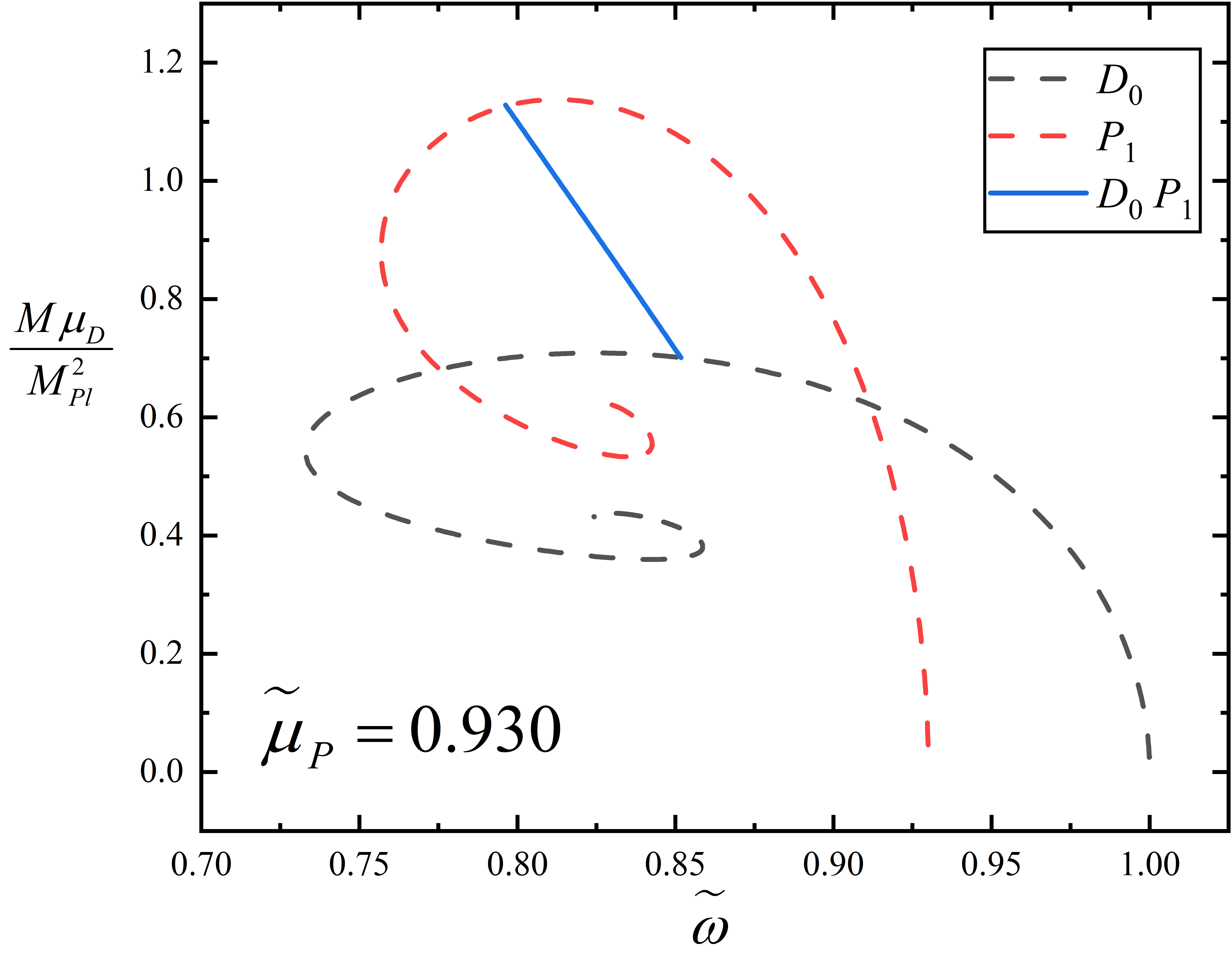}
      \includegraphics[height=.26\textheight]{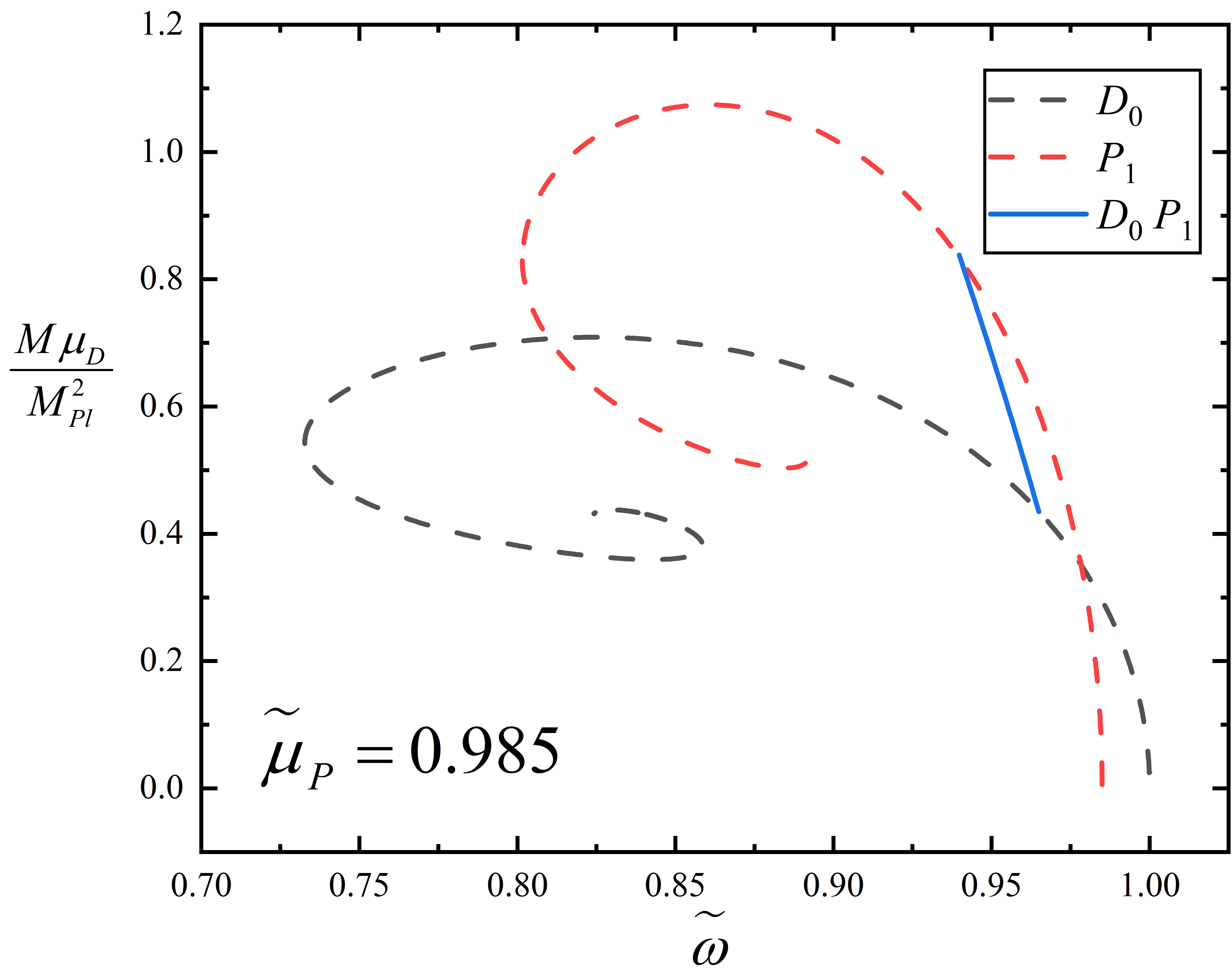}
  \end{center}
  \caption{ The ADM mass $M$ of the DPSs as a function of the synchronized frequency $\tilde{\omega}$ for several values of $\tilde{\mu}_P$.
  }
  \label{ADM2}
\end{figure}

\subsubsection{D-P family}
Next, we  study the \textit{D-P} solutions. This family of solutions is much simpler than the \textit{P-P} family discussed above. The left and right ends of the \textit{D-P} solutions intersect with the Proca field curve and the Dirac field curve, respectively. That is, when the synchronized frequency $\tilde{\omega}$ takes the minimum value, the Dirac field disappears, and the multi-field solution becomes a single-field solution with only Proca field components; when the synchronized frequency $\tilde{\omega}$ takes the maximum value, the Proca field disappears, and the multi-field solution becomes a single-field solution with only Dirac field components. Moreover, compared with the \textit{P-P} solution, there is no distortion in the the ADM mass curve of the \textit{D-P} family.

Fig.~\ref{field2} shows the field functions of the \textit{D-P} solutions at different synchronized frequencies $\tilde{\omega}$. The top two graphs are Dirac field functions ($f$ and $g$), and the bottom two graphs are Proca field functions ($F$ and $G$). For \textit{D-P} family, $\left\lvert f_{max}\right\rvert $ and $\left\lvert g_{max}\right\rvert $ increase monotonically with $\tilde{\omega}$, and $\left\lvert F_{max}\right\rvert $ and $\left\lvert G_{max}\right\rvert $ decrease monotonically with $\tilde{\omega}$. Due to this characteristic of the \textit{D-P} solution, they cannot keep a certain field from disappearing as the synchronized frequency $\tilde{\omega}$ changes.

Fig.~\ref{ADM2} shows the relationship between the ADM mass of different \textit{D-P} solutions and the synchronized frequency $\tilde{\omega}$. Similar to the \textit{P-P} solution, different solutions here are still obtained by changing the parameter $\tilde{\mu}_P$. The \textit{D-P} family is similar to the single-branch solution in Ref.~\cite{Liang:2022mjo}. With increasing $\tilde{\mu}_P$, the \textit{D-P} solutions  gradually move to the right, and its existence domain  gradually become smaller until no solution can be found.

\subsection{ Nonsynchronized frequency }
In this section, we  discuss the non-synchronized frequency solutions $(\tilde{\omega}_D \neq \tilde{\omega}_P)$. In order to study the influence of parameter changes on the properties of the solution, we fix $\tilde{\mu}_D=\tilde{\mu}_P=1$ and obtain a series of solutions by changing $\tilde{\omega}_D$. Similar to the synchronized frequency solutions in the previous section, we still divide it into two categories: \textit{P-P} family and \textit{D-P} family. For $0.726\leq \tilde{\omega}_D \leq 0.743$, the DPS solutions belong to the \textit{P-P} family; for $0.744 < \tilde{\omega}_D \leq 0.980$, the DPS solutions belong to the \textit{D-P} family. Then, we discuss in detail the properties of these two families of solutions.

\subsubsection{P-P family}
\begin{figure}[!htbp]
  \begin{center}
    \includegraphics[height=.26\textheight]{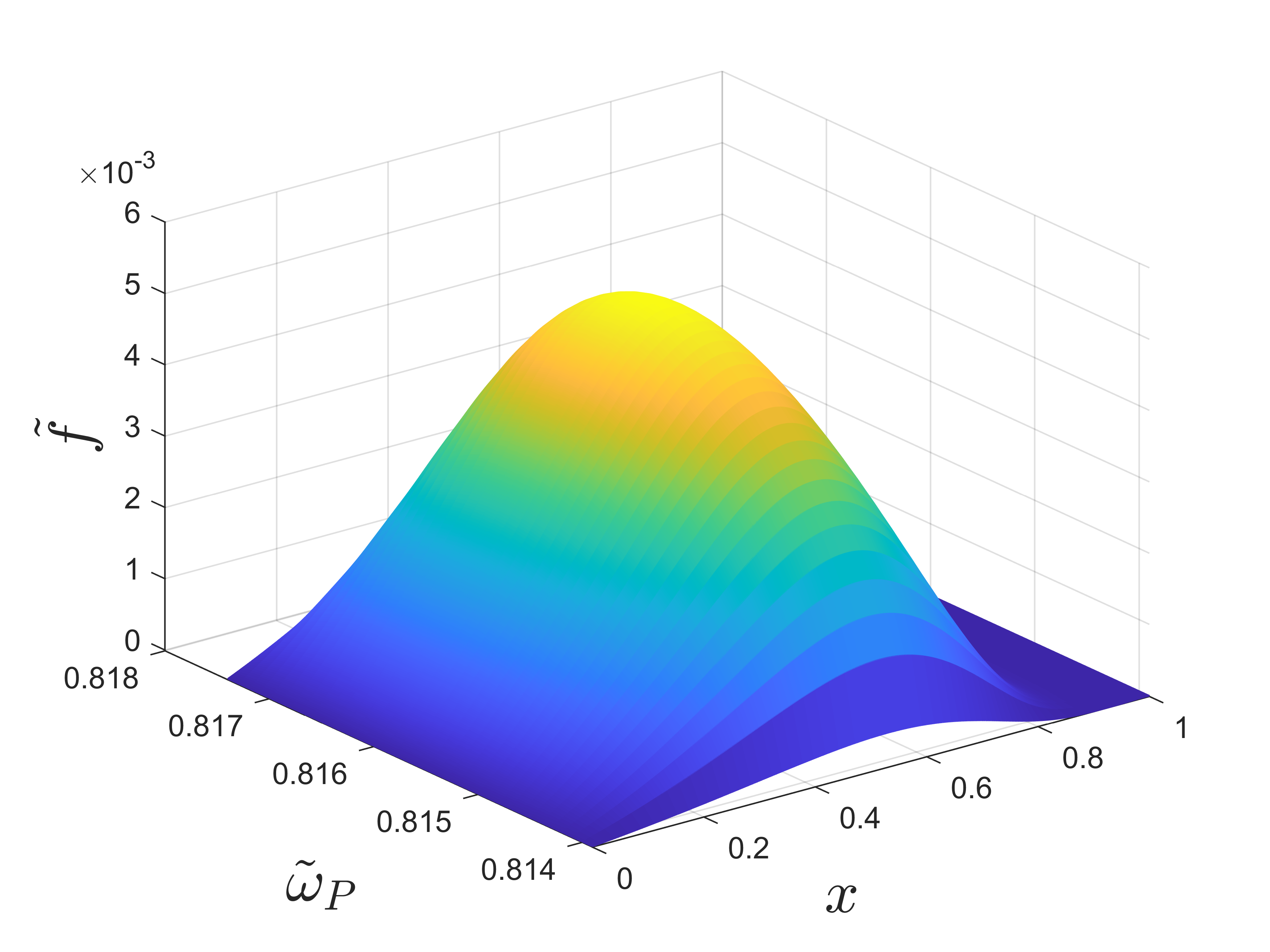}
    \includegraphics[height=.26\textheight]{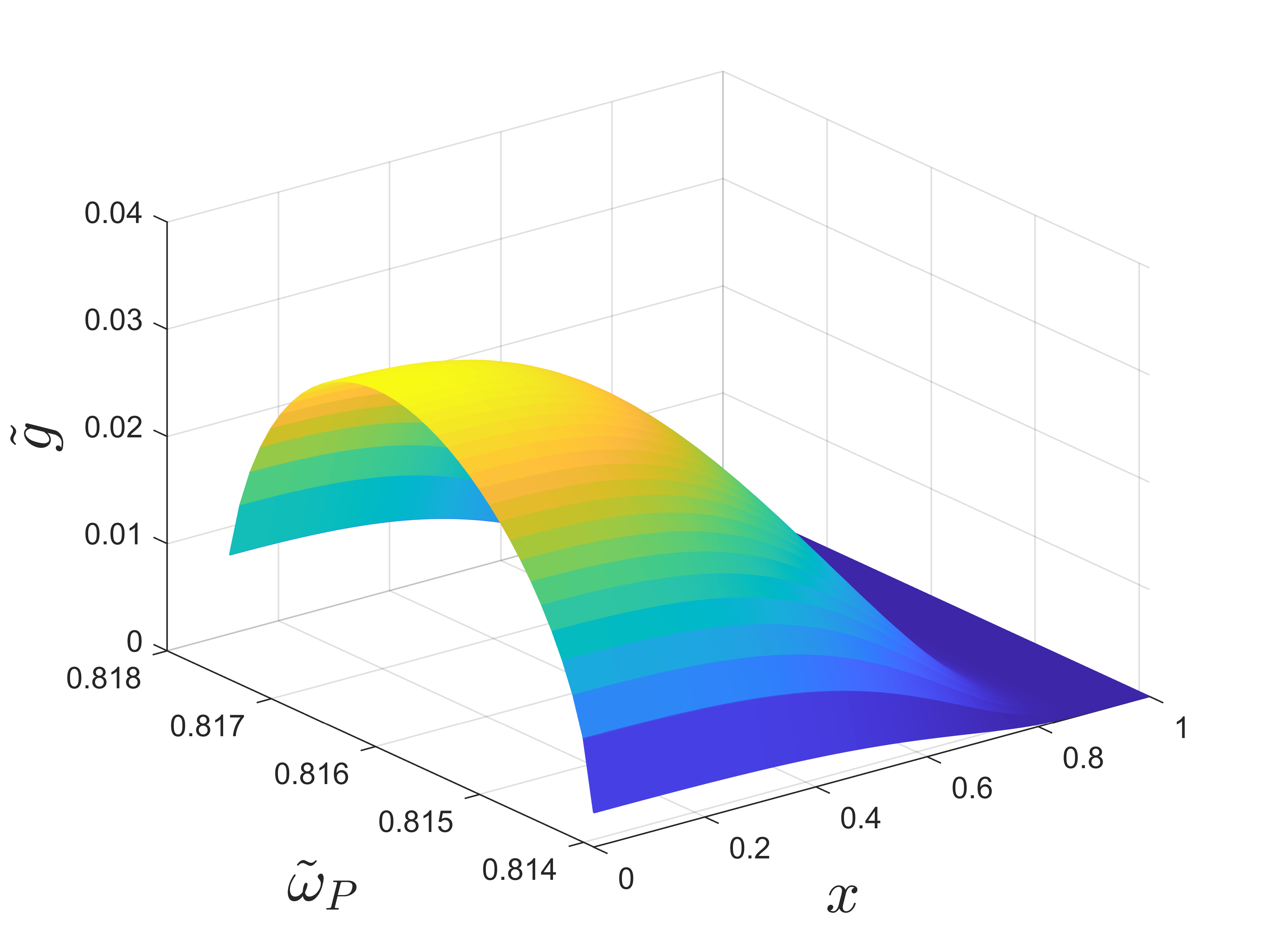}
    \includegraphics[height=.26\textheight]{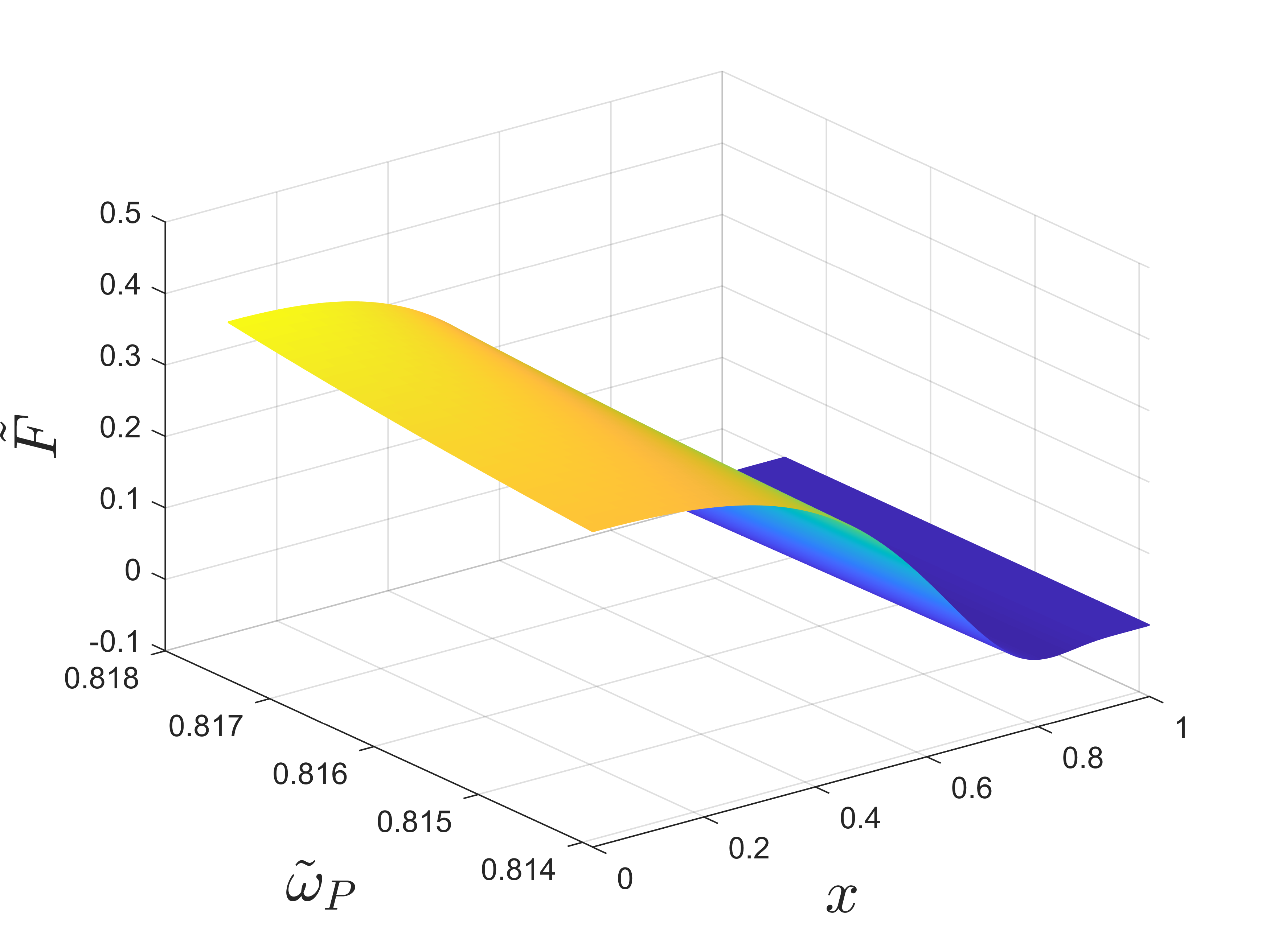}
    \includegraphics[height=.26\textheight]{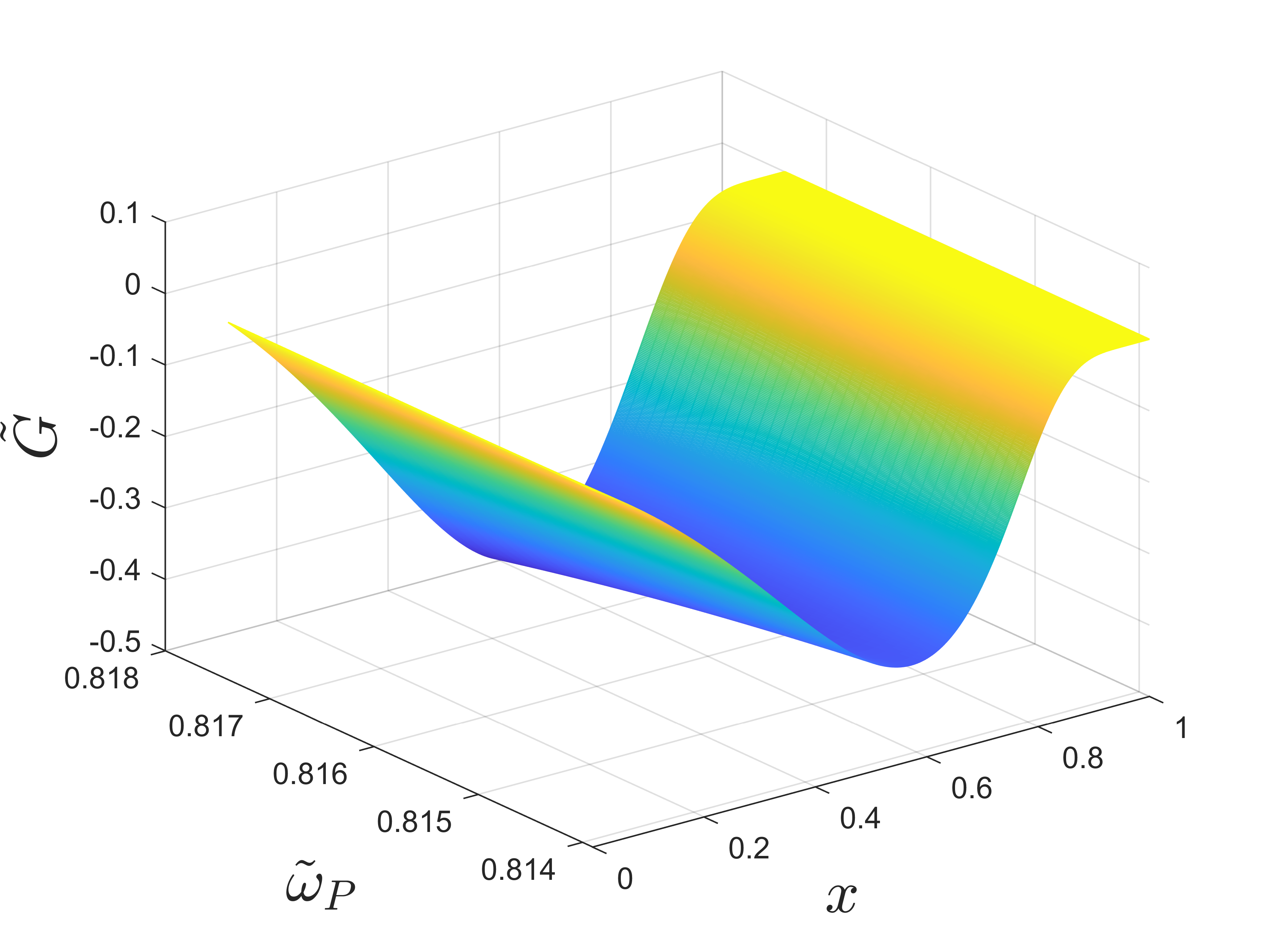}
  \end{center}
  \caption{The matter functions $\tilde{f}$, $\tilde{g}$, $\tilde{F}$ and $\tilde{G}$ as functions of $x$ and $\tilde{\omega}_P$ for $\tilde{\omega}_D = 0.726$.}
  \label{field3}
  \end{figure}

  \begin{figure}[!htbp]
  \begin{center}
    \includegraphics[height=.26\textheight]{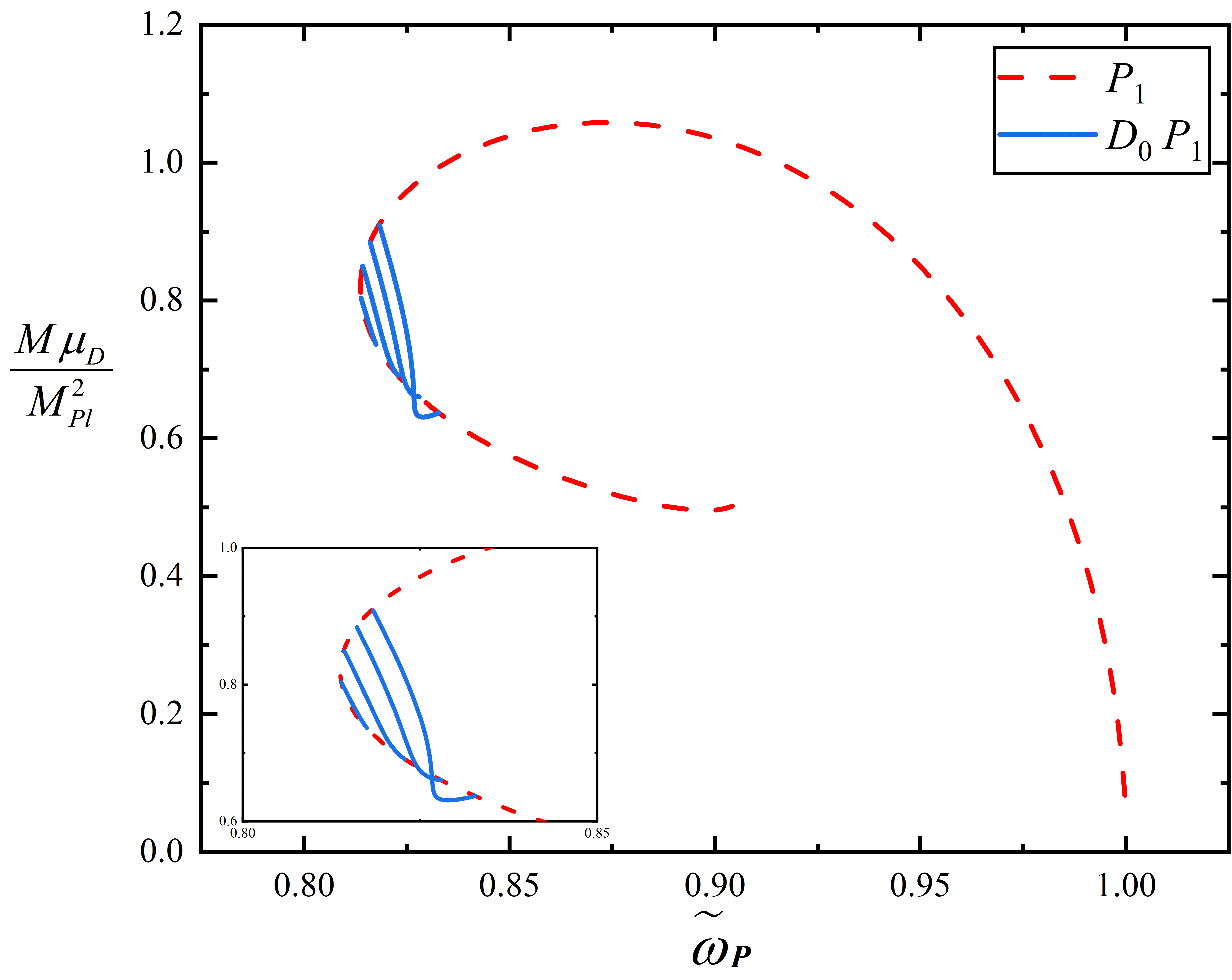}
    \includegraphics[height=.26\textheight]{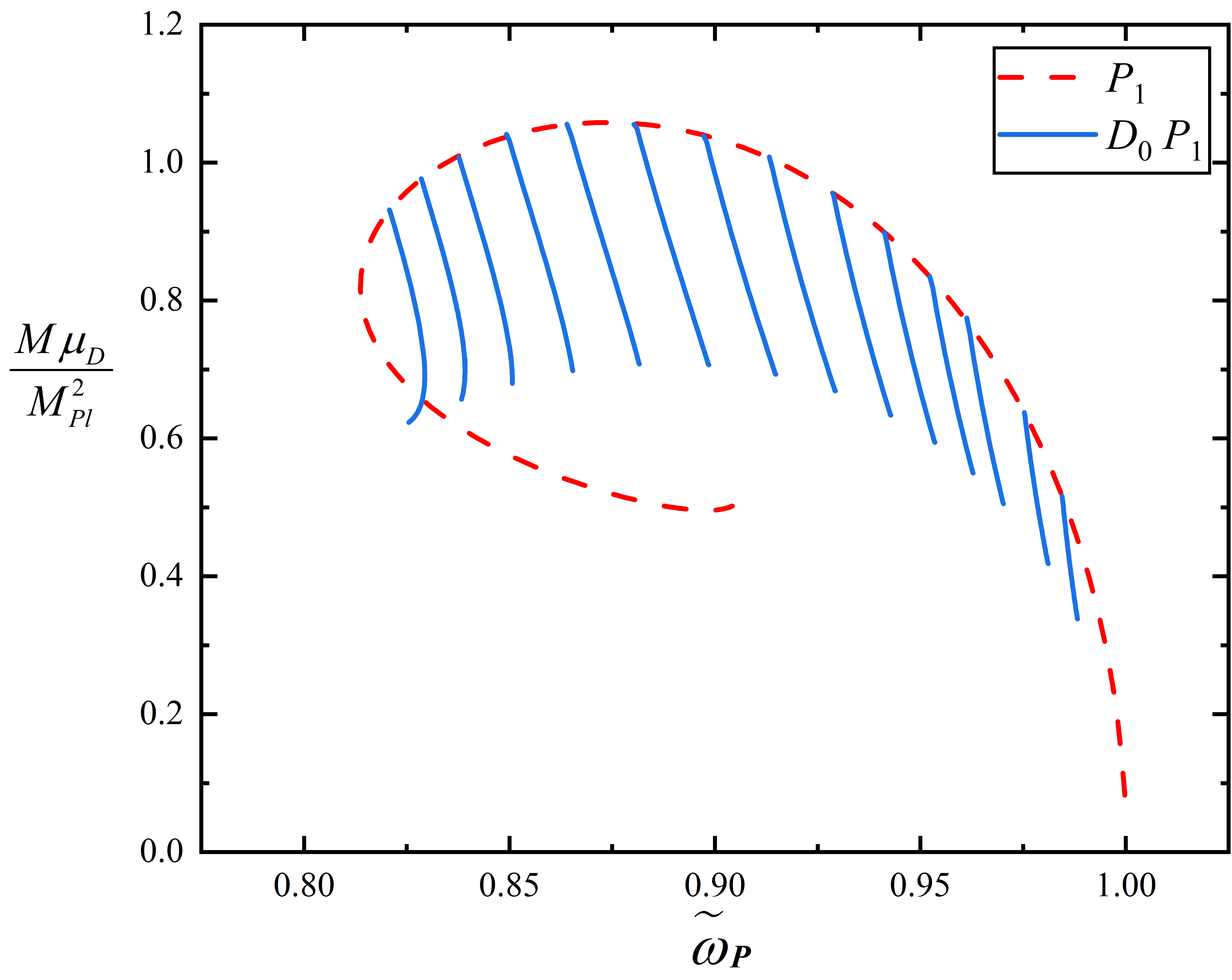}
  \end{center}
  \caption{The ADM mass $M$ of the DPSs as a function of the frequency $\tilde{\omega}_P$ for several values of $\tilde{\omega}_D$. Left: \textit{P-P} family. Right: \textit{D-P} family.}
  \label{ADMnon}
\end{figure}
Fig.~\ref{field3} shows the field functions of different \textit{P-P} solutions at different non-synchronized frequencies $\tilde{\omega}_P$. It can be seen that the behavior of the field functions of the \textit{P-P} solutions under non-synchronized frequency is the same as that under synchronized frequency. $\left\lvert f_{max}\right\rvert $ and $\left\lvert g_{max}\right\rvert $ increase first and then decrease with $\tilde{\omega}_P$, while for the Proca field, $\left\lvert F_{max}\right\rvert $ and $\left\lvert G_{max}\right\rvert $ increase monotonically with $\tilde{\omega}_P$.

Fig.~\ref{ADMnon} represents the ADM mass of the \textit{P-P} family with $\tilde{\omega}_P$. Similar to the \textit{P-P} family under synchronized frequency, the multi-field solutions start from a point on the Proca field single-field curve and finally fall back to another point on the Proca field curve. With increasing $\tilde{\omega}_D$, the \textit{P-P} solutions gradually moves to the right and its existence domain becomes larger.

\subsubsection{D-P family}
\begin{figure}[!htbp]
  \begin{center}
    \includegraphics[height=.26\textheight]{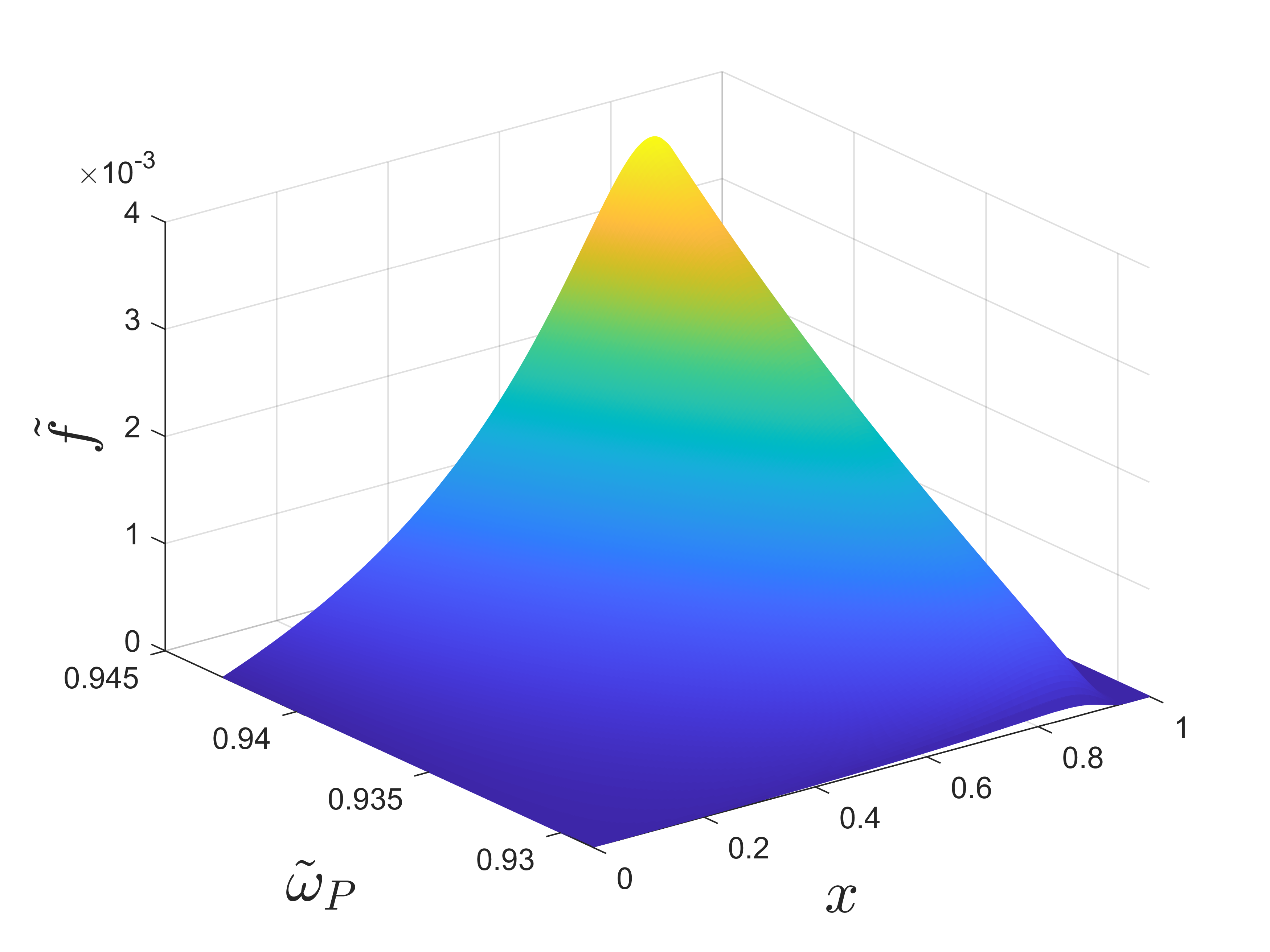}
    \includegraphics[height=.26\textheight]{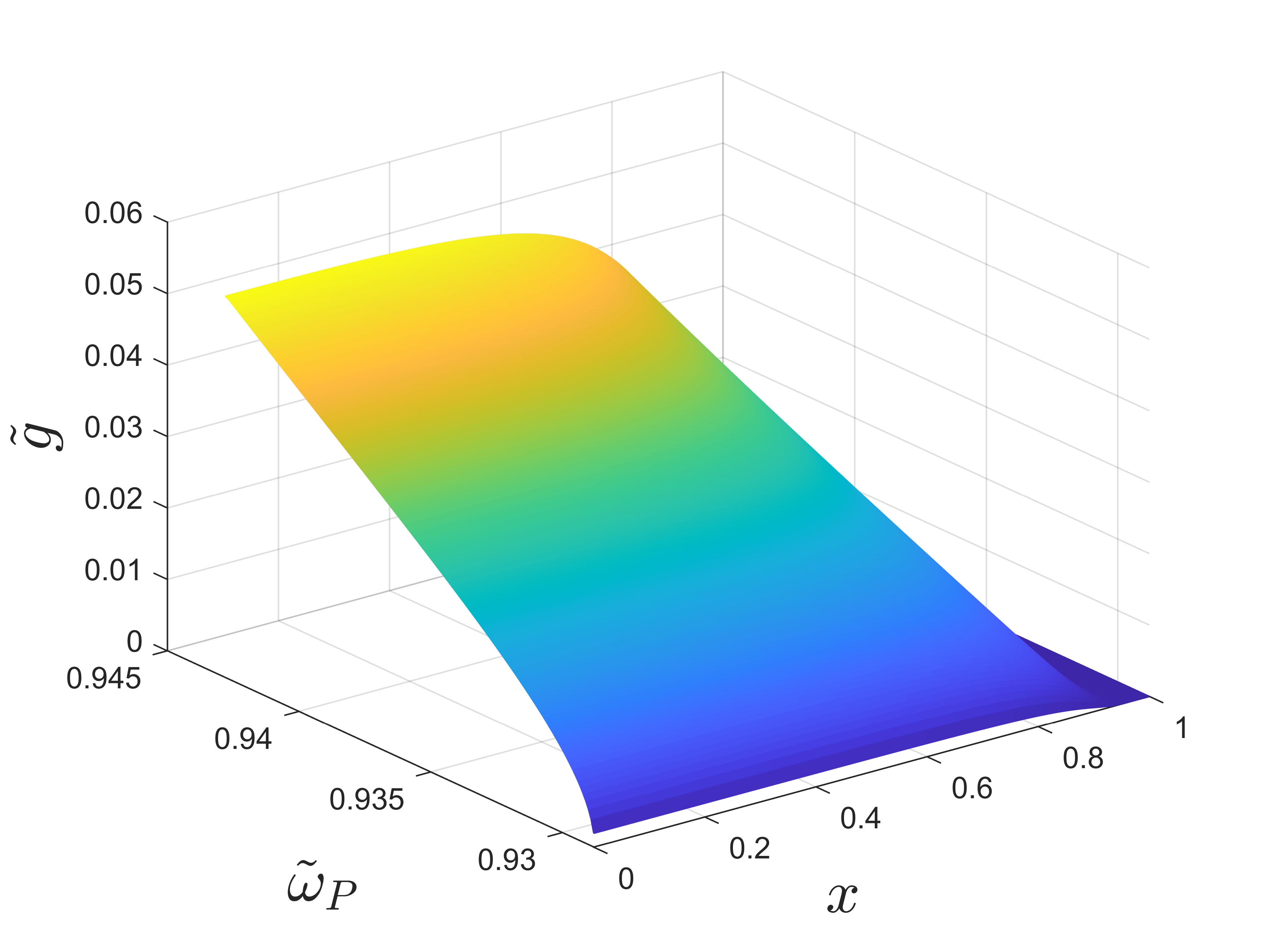}
    \includegraphics[height=.26\textheight]{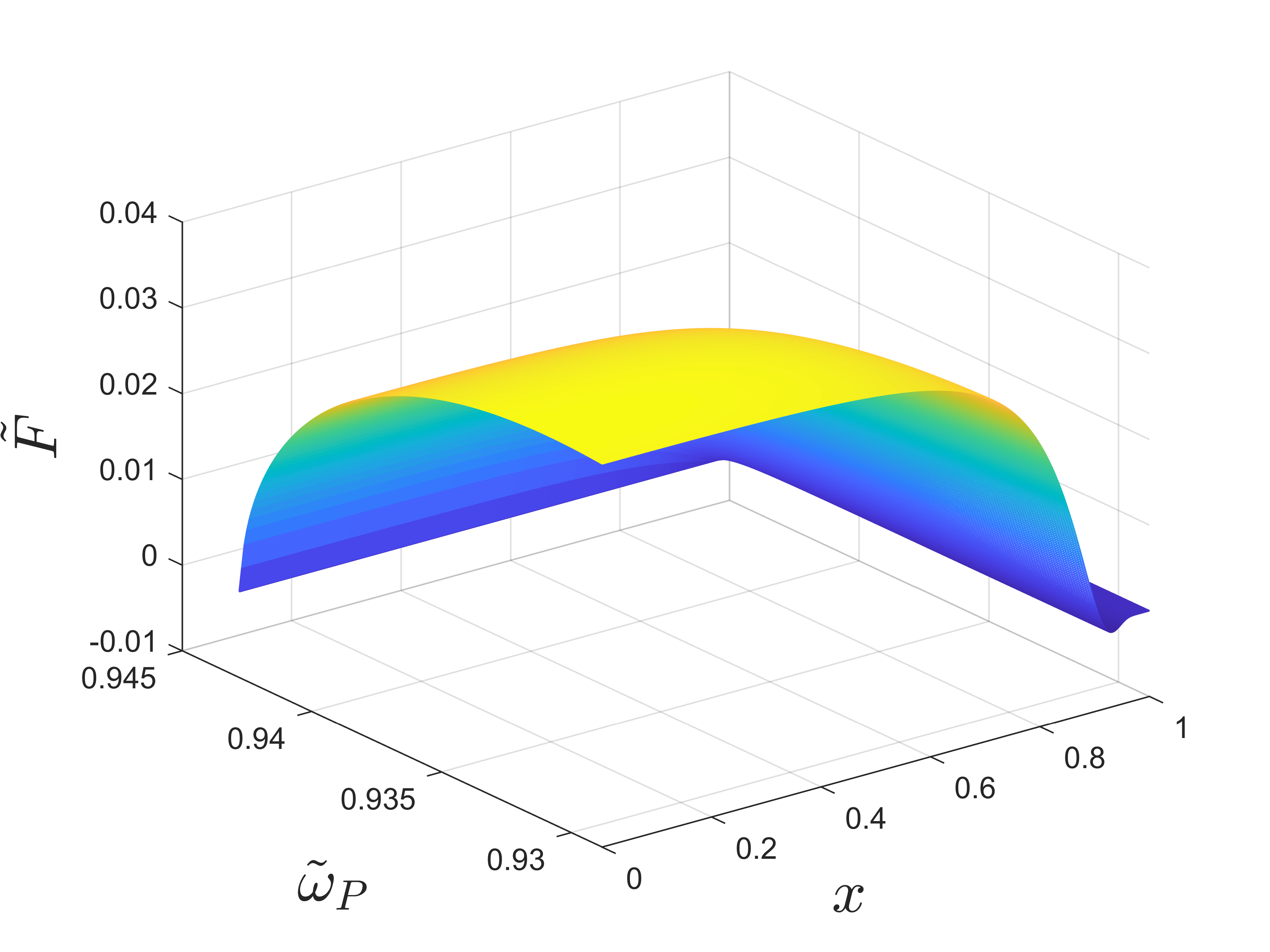}
    \includegraphics[height=.26\textheight]{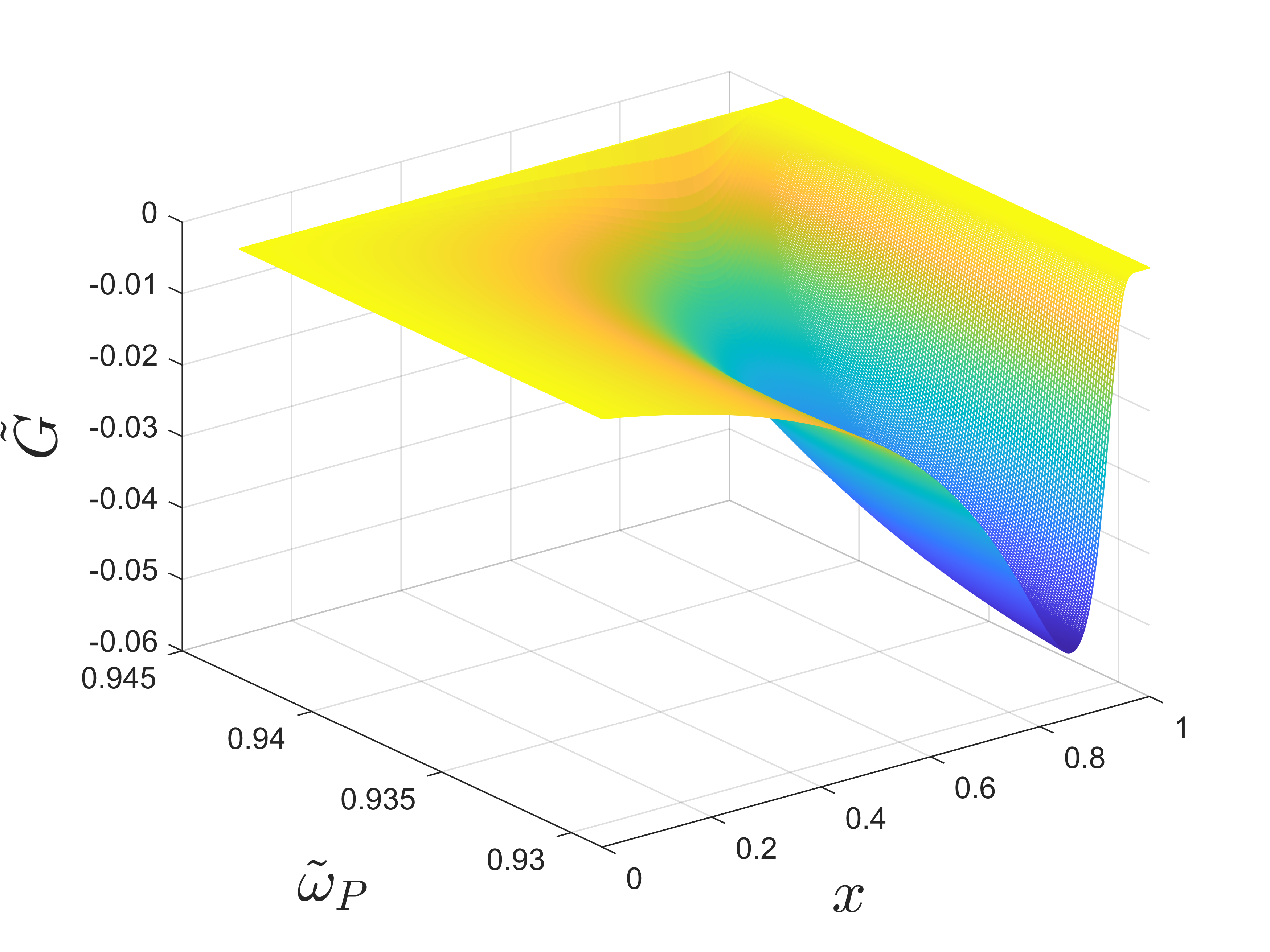}
  \end{center}
  \caption{The matter functions $\tilde{f}$, $\tilde{g}$, $\tilde{F}$ and $\tilde{G}$ as functions of $x$ and $\tilde{\omega}_P$ for $\tilde{\omega}_D = 0.906$.}
  \label{field4}
\end{figure}

For the \textit{D-P} solutions in Fig.~\ref{field4}, its field function evolve behavior is the same as that of the \textit{D-P} solutions under synchronized frequency. When the non-synchronized frequency $\tilde{\omega}_P$ increases, $\left\lvert f_{max}\right\rvert $ and $\left\lvert g_{max}\right\rvert $ increase monotonically, and $\left\lvert F_{max}\right\rvert $ and $\left\lvert G_{max}\right\rvert $ decrease monotonically.

The ADM mass of the \textit{D-P} family with $\tilde{\omega}_D$ is depicted in Fig.~\ref{ADMnon}. Similar to the \textit{D-P} solutions under synchronized frequency, the multi-field solutions starts from a point on the Proca field single-field curve and finally intersects with the Dirac field curve. With increasing $\tilde{\omega}_D$, the multi-field solutions gradually moves to the right and its existence domain becomes smaller until no solution can be found.

\subsection{Binding energy}
We obtain several types of DPS solutions through numerical calculations. We know that the boson star solutions have stable and unstable branches. By analyzing the binding energy, we can preliminarily determine the stability of the solutions. In this section, we  calculate the binding energy of DPS and its ADM mass $M$ at asymptotic infinity in this model. The corresponding Noether charge of Dirac field and Proca field are $Q_S$ and $Q_P$, respectively. The binding energy can be written as
\begin{equation}
  E_B = M -  2 \mu_D Q_D - \mu_P Q_P \,,
\end{equation}
\begin{figure}[!htbp]
  \begin{center}
      \includegraphics[height=.26\textheight]{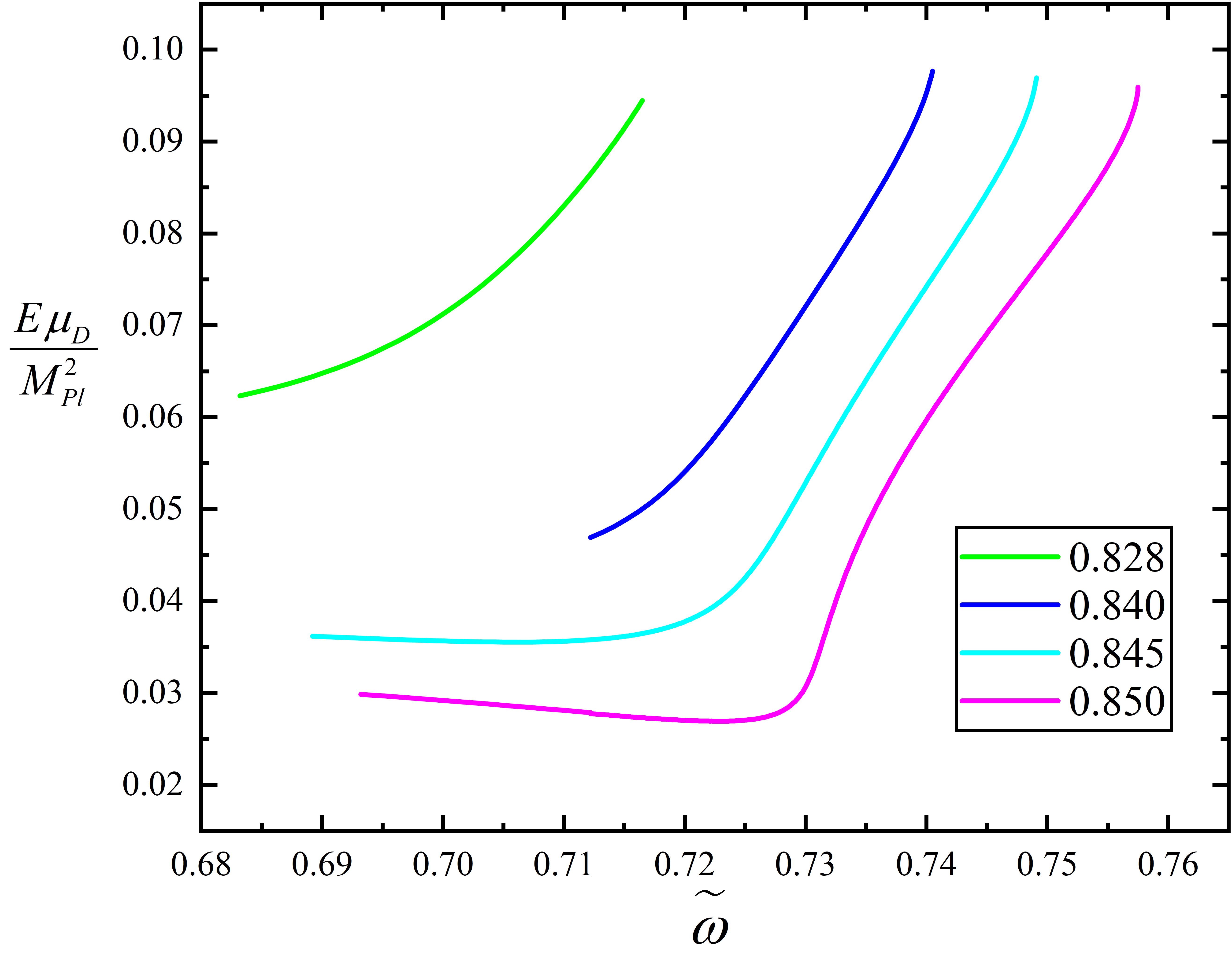}
      \includegraphics[height=.26\textheight]{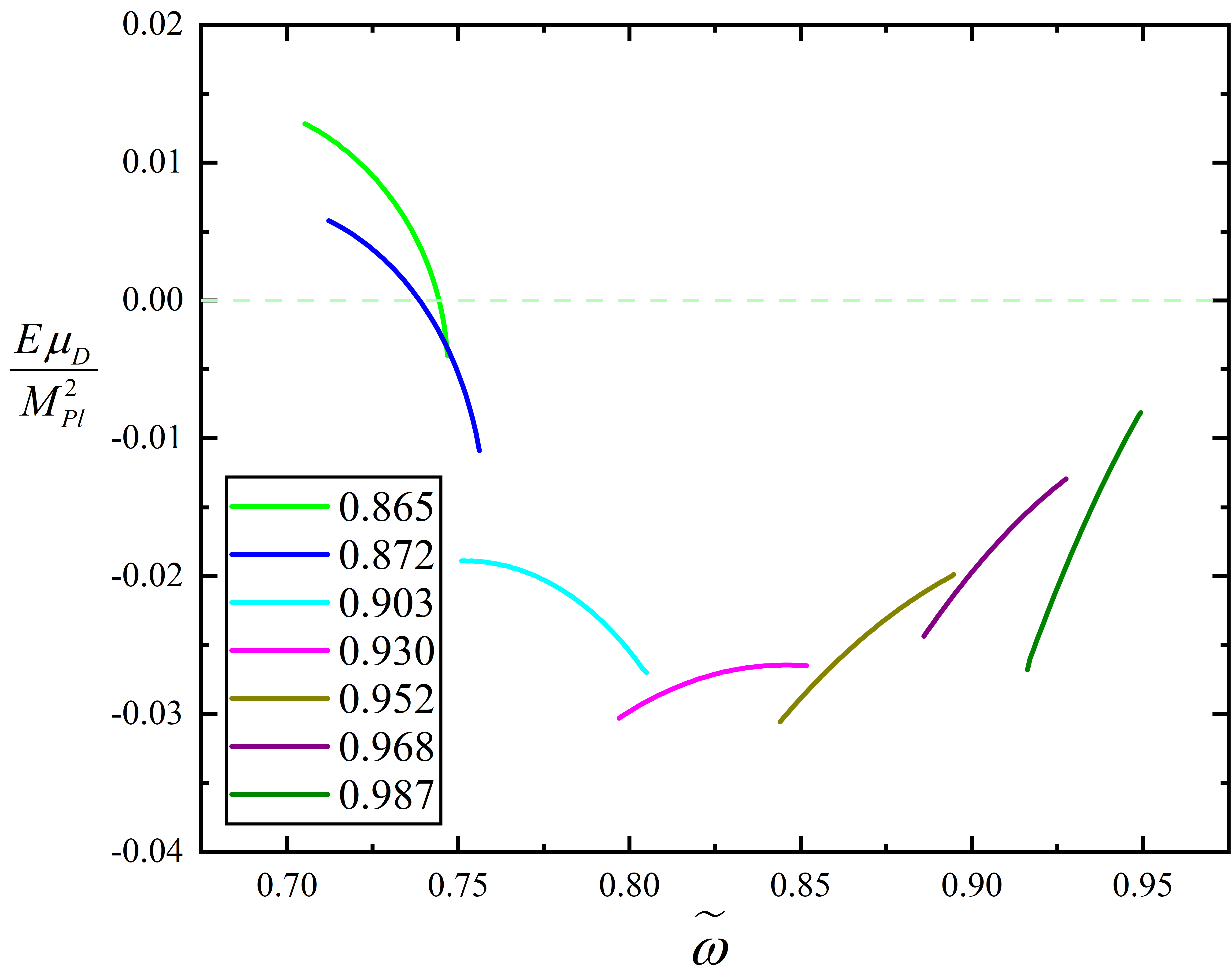}
  \end{center}
  \caption{The binding energy $E$ of the DPSs as a function of the synchronized frequency $\tilde{\omega}$ for several values of $\tilde{\mu}_P$. Left: \textit{P-P} family. Right: \textit{D-P} family.}
  \label{energy1}
\end{figure}

  \begin{figure}[!htbp]
  \begin{center}
      \includegraphics[height=.3\textheight]{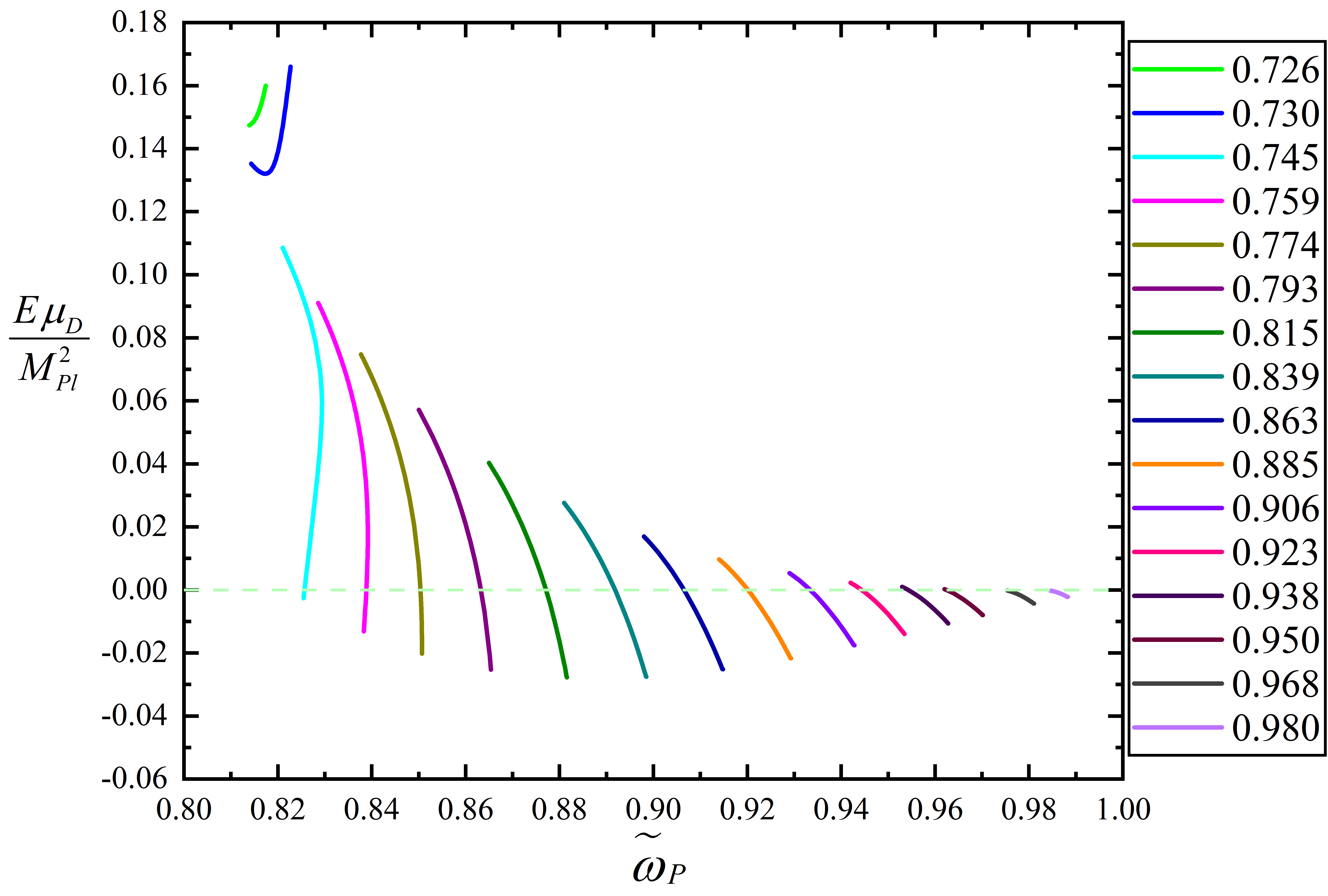}
  \end{center}
  \caption{The binding energy $E$ of the DPSs as a function of the nonsynchronized frequency $\tilde{\omega}_P$ for several values of $\tilde{\omega}_D$.}
  \label{energy2}
\end{figure}

First, let's analyze the binding energy of the synchronized frequency solution. Fig.~\ref{energy1} shows the relationship between the binding energy $E$ and the synchronized frequency $\tilde{\omega}$ in the synchronized frequency case. From the Fig.~\ref{energy1}, it can be concluded that the binding energy of most \textit{D-P} solutions  increase with increasing synchronized frequency $\tilde{\omega}$, and all \textit{D-P} solutions are unstable. However, the behavior of \textit{P-P} solutions is not fixed and most of them are stable. Only when $\tilde{\mu}_P$ is small, they are unstable in a certain frequency domain; conversely, when $\tilde{\mu}_P$ is large, they are all stable within a certain domain.

Finally, let's discuss the non-synchronized frequency case. The relationship between the binding energy of non-synchronized frequency solutions and $\tilde{\omega}_P$ is shown in Fig.~\ref{energy2}. The behavior of binding energy of non-synchronized frequency solutions is relatively trivial. All \textit{P-P} solutions are located in the unstable interval, and the binding energy decreases first and then increases with $\tilde{\omega}_P$. Different $\tilde{\omega}_D$ corresponds to \textit{D-P} solutions with unstable domains in small $\tilde{\omega}_P$ frequency domain and stable domains in large $\tilde{\omega}_P$ frequency domain. There is a critical transition from an unstable to a stable state for all \textit{D-P} solutions.

\section{Conclusion}\label{sec5}
In this article, we have studied the solutions of the Einstein-Dirac-Proca equation by numerical methods, that is, the properties of the spherically symmetric Dirac-Proca star model composed of Dirac field and Proca field. We classified and discussed the field configurations and properties of each component of the Dirac-Proca star under synchronized frequency and non-synchronized frequency conditions, ADM mass, and preliminarily determined the stability of the solutions by discussing the binding energy of each solution.

In this class of solutions under synchronized frequency, we searched for different solutions by changing the value of the Proca field mass $\tilde{\mu}_P$. According to the behavior of the solutions when the synchronized frequency takes extreme values, they are divided into \textit{P-P} family and \textit{D-P} family. When $\tilde{\mu}_P$ is small, the multi-field solutions show that both ends are connected to the Proca field curve in the ADM mass graph. With increasing $\tilde{\mu}_P$, the ADM mass curve in the middle part gradually becomes a multi-value function of $\tilde{\omega}$, but it still belongs to \textit{P-P} family. When $\tilde{\mu}_P$ continues to increase, the ADM mass curve of the multi-field solutions no longer fall on the Proca field curve on the right end but fall on the ADM mass curve of Dirac field. At this time, the multi-field solutions transitions to \textit{D-P} solutions. In addition, we found that the existence domain of \textit{P-P} solutions increases with increasing $\tilde{\mu}_P$, while that of \textit{D-P} solutions decreases with increasing $\tilde{\mu}_P$. When $\tilde{\mu}_P$ is close to 1, \textit{D-P} solutions almost disappear.

For the non-synchronized frequency solution, we fix $\tilde{\mu}_D=\tilde{\mu}_P=1$ and change the Dirac field frequency $\tilde{\omega}_D$ to obtain different solutions. Similar to the synchronized frequency case, we still divide these solutions into two categories: \textit{P-P} family and \textit{D-P} family. As $\tilde{\omega}_D$ gradually increases, the non-synchronized frequency solutions transitions from \textit{P-P} to \textit{D-P}. When $\tilde{\omega}_D$ reaches its maximum value, the existence domain becomes very narrow and the multi-field solutions almost disappears. In \textit{P-P} solutions, the Proca field component never disappears, while in \textit{D-P} solutions, both field functions corresponding to the two fields change monotonically and one field disappears at one end and the other field disappears at the other end.

Regarding the stability of DPS, we calculated the relationship between binding energy and synchronized frequency (non-synchronized frequency). From the analysis we concluded that all \textit{P-P} solutions are unstable. Some \textit{D-P} solutions corresponding to certain $\tilde{\mu}_P$ values in synchronized frequency may be stable in all existence domains, and other solutions are stable only in some domains; The binding energy behavior of \textit{D-P} solutions under non-synchronized frequency is consistent, and they are only stable in some domains.

It is worth noting that the behavior of the ADM mass curve in this article is very different from those in previous studies on multi-field boson star solutions. Generally speaking, the number of branches of multi-field solutions corresponds one-to-one with the behavior of the solutions at both ends. However, in the DPS model, the above conditions are not satisfied, and the ADM mass function, which is shaped like a spoon, also undergoes distortion. As for the cause of this distortion, we have not fully understood it yet.
\section*{Acknowledgements}
This work is supported by National Key Research and Development Program of China (Grant No.~2020YFC2201503) and  the National Natural Science Foundation of China (Grants No.~12275110 and No.~12247101). Parts of computations were performed on the shared memory system at institute of computational physics and complex systems in Lanzhou university.

\providecommand{\href}[2]{#2}\begingroup\raggedright

  \end{document}